# THERMODYNAMICS OF 2-ALKANOL + POLAR ORGANIC SOLVENT MIXTURES. I. SYSTEMS WITH KETONES, ETHERS OR ORGANIC CARBONATES

Juan Antonio González,[*] Fernando Hevia, Luis Felipe Sanz, Daniel Lozano-Martín and Isaías García de la Fuente

[a]G.E.T.E.F., Departamento de Física Aplicada, Facultad de Ciencias, Universidad de Valladolid, Paseo de Belén, 7, 47011 Valladolid, Spain.

*corresponding author, e-mail: jagl@termo.uva.es; Fax: +34-983-423136; Tel: +34-983-423757

**Abstract**

The mixtures 2-propanol or 2-butanol + *n*-alkanone, or + acetophenone, or + linear monoether, or + cyclic ether, or + linear organic carbonate, or + propylene carbonate have been investigated using thermodynamic data, and in terms of the Flory theory, and the Kirkwood-Buff integrals. The data considered are: excess molar enthalpies ($H_m^E$), volumes, entropies, and the temperature dependence of $H_m^E$. The enthalpy of the 2-alkanol-solvent interactions have been determined, and the different contributions to $H_m^E$ discussed. It is shown that $H_m^E$ values of the 2-alkanol (fixed) + *n*-alkanone, or + linear carbonate mixtures change in the same manner that for *n*-alkanone, or linear carbonate + *n*-alkane (fixed) systems. In contrast, $H_m^E$ values of 2-alkanol (fixed) + linear monoether or + *n*-alkane mixtures change similarly. This set of results suggests that solvent-solvent interactions are determinant in systems with *n*-alkanone or linear carbonate, while interactions between alcohol molecules are determinant in mixtures with linear monoethers. According to the Flory model, orientational effects in systems with a given 2-alkanol become weaker in the sequence: linear monoether > linear organic carbonate > *n*-alkanone, and are stronger in solutions with a cyclic monoether than in those with cyclic diethers, and in systems with acetophenone or propylene carbonate than in the mixtures with the corresponding linear solvents. Results obtained from the Kirkwood-Buff integrals are consistent with these findings. The application of Flory model reveals that orientational effects are similar in systems with 1- or 2-alkanols, with the exception of solutions with linear monoethers, where such effects are stronger in mixtures containing 1-alkanols. For the examined systems, the formalism of the Kirkwood-Buff integrals makes no meaningful distinction between solutions with 1-alkanols or 2-alkanols.

**Keywords**: 2-alkanol; polar solvent; thermodynamic data; Flory; Kirkwood-Buff; orientational effects

## 1. Introduction

Hydrogen bonding is fundamental in basic sciences and in industrial applications. In addition, this attractive interaction together with the C-C interactions are essential for humans since water, a H-bonding compound, is crucial for the maintenance and development of life. It is well-known that alcohols are self-associated molecules which form molecular aggregates when are mixed with inert and non-polar solvents. The size and shape (linear, cyclic) of the formed clusters can be investigated using, e.g., thermodynamic data [1-5], or a variety of different spectroscopic techniques [6-10] or dielectric measurements [11-16]. On the other hand, experimental data have been employed to investigate this type of solutions by means of different models: association theories, [4,17,18], group contribution models [19,20], equations of state [21], the Kirkwood correlation factor [12,18], or the Kirkwood-Buff integrals [22,23], amongst others.

We have studied changes produced in the alcohol network and orientational effects in systems with 1-alkanol and ether [24,25], or $n$-alkanone [26], or linear organic carbonate [27], or + nitrile [28] on the basis of the thermodynamic data available, and using different models such as: Flory [29], the Kirkwood-Buff integrals, or the formalism of the concentration-concentration structure factor [30]. The main conclusion of these studies is that random mixing is a good approximation for describing the behavior of many of the mentioned solutions. As a continuation of these works, we pay now attention to 2-alkanol + polar organic solvent mixtures. In the present research the solvents considered are: $n$-alkanones (propanone, 2-butanone, 2-pentanone, 3-pentanone), acetophenone, linear monoethers (dipropyl ether (DPE) and dibutyl ether (DBE)), cyclic ethers (oxolane (THF), oxane (THP), 1,4-dioxane, 1,3-dioxolane) and organic carbonates (dimethyl carbonate (DMC), diethyl carbonate (DEC), or propylene carbonate (PC)). The research is conducted using experimental data from the literature such as excess molar enthalpies ($H_m^E$), volumes ($V_m^E$), or entropies ($TS_m^E = H_m^E - G_m^E$; $G_m^E$, molar excess Gibbs energy), and by means of the application of the Flory model and the Kirkwood-Buff integrals. A comparison with the orientational effects present in systems with the corresponding 1-alkanols is also shown. Many of these results were taken from our previous studies on solutions with 1-alkanols [24,26,27]. For the sake of completeness, we have also conducted the needed calculations for 1-propanol, or 1-butanol + acetophenone, or + cyclic diether mixtures, since they have not been previously investigated using the Flory model. Finally, it should be mentioned that different association theories, including a physical term, have been applied to the study of some systems formed by 2-propanol, or 2-butanol and a polar organic solvent (see, e.g, [31-34]).

## 2. Theories

### 2.1 Flory model

The main features of the theory are the following [29,35,36]. (i) Molecules are divided into segments, which is an arbitrarily chosen isomeric portion of the molecule. The number of segments per molecule of component $i$ is denoted by $r_i$ and the number of intermolecular contact sites per segment by $s_i$. (ii) The mean intermolecular energy per contact is proportional to $-\eta/v_s$ (where $\eta$ is a positive constant characterizing the energy of interaction for a pair of neighbouring sites and $v_s$ is the volume of a segment). (iii) The configurational partition function is stated under the assumption than the number of external degrees of freedom of the segments is < 3. This is needed to take into account the restrictions on the precise location of a given segment by its neighbours in the same chain. (iv) Random mixing is assumed: the probability of having species of kind $i$ neighbours to any given site is equal to the site fraction, $\theta_i$. For very large total number of contact sites, the probability of formation of an interaction between contacts sites belonging to different liquids is $\theta_1\theta_2$. Under these hypotheses, the Flory equation of state is:

$$\frac{\hat{P}\hat{V}}{\hat{T}} = \frac{\hat{V}^{1/3}}{\hat{V}^{1/3}-1} - \frac{1}{\hat{V}\hat{T}} \tag{1}$$

where $\hat{V} = V_m/V_m^*$; $\hat{P} = P/P^*$ and $\hat{T} = T/T^*$ are the reduced volume, pressure and temperature, respectively ($V_m$ is the molar volume of the mixture). Equation (1) is valid for pure liquids and liquid mixtures. For pure liquids, the reduction parameters, $V_{mi}^*$, $P_i^*$ and $T_i^*$ can be obtained from experimental data, such as density, $\alpha_{pi}$ (isobaric expansion coefficient) and $\kappa_{Ti}$ (isothermal compressibility). For mixtures, the corresponding parameters are calculated as follows:

$$V_m^* = x_1 V_{m1}^* + x_2 V_{m2}^* \tag{2}$$

$$T^* = \frac{\varphi_1 P_1^* + \varphi_2 P_2^* - \varphi_1 \theta_2 X_{12}}{\dfrac{\varphi_1 P_1^*}{T_1^*} + \dfrac{\varphi_2 P_2^*}{T_2^*}} \tag{3}$$

$$P^* = \varphi_1 P_1^* + \varphi_2 P_2^* - \varphi_1 \theta_2 X_{12} \tag{4}$$

In equatios. (3) and (4), $\varphi_i = x_i V_{mi}^* / \sum x_i V_{mi}^*$ is the segment fraction and $\theta_2 = \varphi_2 /(\varphi_2 + S_{12}\varphi_1)$. $S_{12}$ is the geometrical parameter of the mixture, which, in the case of spherical molecules, is calculated as $S_{12} = (\frac{V_{m1}^*}{V_{m2}^*})^{-1/3}$. The energetic parameter, $X_{12}$, also present in equations (3) and (4), is defined, by similarity with $P_i^* = \frac{s_i \eta_{ii}}{2 v_s^{*2}}$, as

$$X_{12} = \frac{s_1 \Delta \eta}{2 v_s^{*2}} \tag{5}$$

where $\Delta \eta = \eta_{11} + \eta_{22} - 2\eta_{12}$. In equation (5), $v_s^*$ (reduction volume for segment) and $\eta_{ij}$ are changed from molecular units to molar units per segments. $X_{12}$ is determined from:

$$H_m^E = \frac{x_1 V_{m1}^* \theta_2 X_{12}}{\hat{V}} + x_1 V_{m1}^* P_1^* (\frac{1}{\hat{V}_{m1}} - \frac{1}{\hat{V}}) + x_2 V_{m2}^* P_2^* (\frac{1}{\hat{V}_{m2}} - \frac{1}{\hat{V}}) \tag{6}$$

The reduced volume of the mixture, $\hat{V}$, is obtained from the equation of state. Therefore, the molar excess volume can be also calculated:

$$V_m^E = (x_1 V_{m1}^* + x_2 V_{m2}^*)(\hat{V} - \varphi_1 \hat{V}_{m1} - \varphi_2 \hat{V}_{m2}) \tag{7}$$

*2.2   Kirkwood-Buff integrals*

In the framework of this theory [37,38], the Kirkwood-Buff integrals are determined from:

$$G_{ij} = \int_o^\infty (g_{ij} - 1) 4\pi r^2 dr \tag{8}$$

where, $g_{ij}$ is the radial distribution function (probability of finding a molecule of species i in a volume element at the distance r of the center of a molecule of type j). Therefore, $g_{ij}$ provides information about the mixture structure at microscopic level. Values of $G_{ij}$ can be interpreted as follows: positive values represent the excess of molecules of the type i in the space around a given molecule of kind j. That is, attractive interactions between molecules of i and j exist. Negative values of $G_{ij}$ reveal that i-i and j-j interactions are predominant over i-j interactions [37,39]. The Kirkwood-Buff integrals can be derived from thermodynamic data such as chemical potential; partial molar volumes and isothermal compressibility factor. The resulting equations are [37,40]:

$$G_{ii} = RT\kappa_T + \frac{x_j \bar{V}_{mj}^2}{x_i V_m D} - \frac{V_m}{x_i} \quad (i,j=1,2) \tag{9}$$

$$G_{12} = G_{21} = RT\kappa_T - \frac{\bar{V}_{m1}\bar{V}_{m2}}{V_m D} \tag{10}$$

where $R$ is the gas constant, $x_i$ and $\bar{V}_{mi}$ are the mole fraction and the partial molar volume of component i, respectively (i = 1,2) and $\kappa_T$, the isothermal compressibility of the mixture. D is defined as:

$$D = 1 + \frac{x_1 x_2}{RT}\left(\frac{\partial^2 G_m^E}{\partial x_1^2}\right)_{P,T} \tag{11}$$

The $G_{ij}$ integrals allow estimate the linear coefficients of preferential solvation [41]:

$$\delta_{ii}^0 = x_i x_j G_{ii} - G_{ij})$$
$$\delta_{ij}^0 = x_i x_j (G_{ij} - G_{jj}) \tag{12}$$

which are useful quantities to determine the local mole fractions of the i species around the central j molecule [41].

### 3   Model calculations and results

Values of the properties of pure 2-alkanols and of acetophenone, needed for calculations, are listed in Table 1. For the remainder organic solvents considered, the values used have been taken from previous applications: ketones and organic carbonates [27,42], linear monoethers [24], cyclic ethers [43]. For 1-alkanols, the corresponding values are the same as in reference [25].

*3.1   Flory*

Table 2 lists values of the interaction parameter $X_{12}$ determined from $H_m^E$ results at equimolar composition and 298.15 K [44] and the source of data used in this application. In addition, the relative standard deviations for $H_m^E$ defined as:

$$\sigma_r(H_m^E) = \left[\frac{1}{N}\sum\left(\frac{H_{m,exp}^E - H_{m,calc}^E}{H_{m,exp}^E}\right)^2\right]^{1/2} \tag{13}$$

are also given in Table 2 (see Figures 1-4). Figure S1 (supplementary material) shows the concentration dependence of $X_{12}$ for a few systems. Such dependence has been determined according to the method explained in detail elsewhere [24]. Results provided by the model on $V_m^E$ are collected in Table S1.

### 3.2 Kirkwood-Buff integrals

Along this type of calculations, isothermal compressibilities of the mixtures were determined from $\kappa_T = \Phi_1\kappa_{T1} + \Phi_2\kappa_{T2}$ (where $\Phi_i$ is the volume fraction of the component i of the system), i.e., ideal behaviour is assumed). This is a typical approach, which has not influence on the final calculations of the Kirkwood-Buff integrals [45]. Values of $D$ were obtained using Redlich-Kister type expressions for $G_m^E$ determined from vapour-liquid equilibrium (VLE) data at 298.15 K. Results on $\delta_{ij}^0 = x_i x_j (G_{ij} - G_{jj})$ are collected in Table 3 (Figures 5,6), together with the source of experimental data on VLE and $V_m^E$ used.

### 4. Discussion

Below, excess molar properties are given at equimolar composition and 298.15 K. On the other hand, X stands for the characteristic functional group of the considered organic solvent, in this study alkanones, ethers or organic carbonates.

Values of $H_m^E$ of 2-alkanol + organic solvent mixtures listed in Table 2 are positive, indicating that interactions between like molecules are dominant. The corresponding $V_m^E$ values are usually positive and therefore the main contribution to this excess function comes from interactional effects. Negative $V_m^E$ values are encountered for a few systems. For example, $V_m^E$/cm$^3$ mol$^{-1}$ = $-0.247$ (2-propanol + acetophenone) [46]; $-0.027$ (2-propanol + dipropyl ether) [47] or $-0.073$ (2-butanol + dipropyl ether) [31]. In such cases, the main contribution to $V_m^E$ arises from structural effects.

### 4.1 Enthalpies of the hydroxyl-solvent interactions

If structural effects are neglected [48,49], $H_m^E$ can be determined by adding three contributions: two positive, $\Delta H_{OH-OH}, \Delta H_{X-X}$, due, respectively, to the breaking of alkanol-alkanol and solvent-solvent interactions upon mixing, and a third negative contribution,

$\Delta H_{\text{OH-X}}$, related to the new interactions between unlike molecules created along the mixing process. Thus [50-53]:

$$H_{\text{m}}^{\text{E}} = \Delta H_{\text{OH-OH}} + \Delta H_{\text{X-X}} + \Delta H_{\text{OH-X}} \qquad (14)$$

The enthalpy of the H-bonds between 2-alkanols and the considered organic solvents, termed as $\Delta H_{\text{OH-X}}^{\text{int}}$ may be evaluated by extending the equation (14) to $x_1 \to 0$ [53-55]. Then, $\Delta H_{\text{OH-OH}}$ and $\Delta H_{\text{X-X}}$ can be replaced by $H_{\text{m1}}^{\text{E},\infty}$ (partial excess molar enthalpy at infinite dilution of the first component) of 2-alkanol or organic solvent + heptane mixtures, and $H_{\text{m}}^{\text{E}}$ by the corresponding $H_{\text{m1}}^{\text{E},\infty}$ result of 2-alkanol + organic solvent systems. That is,

$$\Delta H_{\text{OH-X}}^{\text{int}} = H_{\text{m1}}^{\text{E},\infty}(2-\text{alkanol} + \text{organic solvent})$$
$$-H_{\text{m1}}^{\text{E},\infty}(2-\text{alkanol} + \text{heptane}) - H_{\text{m1}}^{\text{E},\infty}(\text{organic solvent} + \text{heptane}) \qquad (15)$$

For 2-alkanol + $n$-alkane mixtures, it was assumed that $H_{\text{m1}}^{\text{E},\infty}$ is independent of the alcohol, a typical approach within association theories [17,56-59]. In this work, we have used $H_{\text{m1}}^{\text{E},\infty} = 23.2$ kJ mol$^{-1}$. This is the same value that for mixtures with 1-alkanols [60-62]. The following features support such choice. (i) The direct experimental calorimetric result for the 2-propanol + heptane system at 303.15 K is 23.7 kJ mol$^{-1}$ [63]; for the 2-propanol + cyclohexane mixture at 313.15 K, this value is $H_{\text{m1}}^{\text{E},\infty} = 21.3$ kJ mol$^{-1}$ [64]. From $H_{\text{m}}^{\text{E}}$ measurements at 298.15 K over the whole composition range, the $H_{\text{m1}}^{\text{E},\infty}$ values are: 24.7 kJ mol$^{-1}$ for the systems 2-propanol + heptane, or 2-butanol + hexane [65]. (ii) Applications of the ERAS model to mixtures containing 2-alkanols have been conducted assuming that the enthalpy of hydrogen bonding between 2-alkanols molecules is $-25.1$ kJ mol$^{-1}$, the same value that for systems with 1-alkanols [17,58,59]. Values of $\Delta H_{\text{OH-X}}^{\text{int}}$ are listed in Table 4. We note that, for mixtures involving a solvent characterized by a functional group X, $\Delta H_{\text{OH-X}}^{\text{int}}$ is, in practice, independent of the alcohol, and of the size of the solvent. $\Delta H_{\text{OH-X}}^{\text{int}}$ results allow conclude that interactions between unlike molecules become stronger in the sequence: linear monoether < linear organic carbonate < $n$-alkanone. On the other hand, alkanol-ether interactions are stronger in the case of cyclic ethers (Table 4).

Values $\Delta H_{\text{OH-CO}}^{\text{int}}$ have been determined using different theories. The application of an association model including physical interactions described by the NRTL equation yields $\Delta H_{\text{OH-CO}}^{\text{int}} = -21$ kJ mol$^{-1}$ for the 2-propanol + propanone system [32]. In terms of the UNIQUAC-associated solution model [34,66], $\Delta H_{\text{OH-CO}}^{\text{int}} = -14$ kJ mol$^{-1}$ for the 2-butanol + propanone mixture [33], and $\Delta H_{\text{OH-O}}^{\text{int}} = -16.3$ kJ mol$^{-1}$ for the 2-propanol + 1,4-dioxane system [67]. Accordingly to the ERAS model, $\Delta H_{\text{OH-O}}^{\text{int}} = -18.5$ kJ mol$^{-1}$ for the 2-butanol + dipropyl ether mixture [31]. Values obtained using equilibrium constants at different temperatures determined form diamagnetic susceptibility data are [68]: $\Delta H_{\text{OH-CO}}^{\text{int}}$ / kJ mol$^{-1}$ = $-17.9$ (2-propanol + propanone); $-18.7$ (2-propanol + 2-butanone) and $\Delta H_{\text{OH-O}}^{\text{int}}$ / kJ mol$^{-1}$ = $-18.7$ (2-propanol + THF). Finally, the value of $\Delta H_{\text{OH-CO}}^{\text{int}}$ for the 2-butanol + propanone mixture at 303.15 K determined by a thermochemical cycle, which takes into account the dipolar stabilitation energy of the 2-butanol-acetone complex, is $-17.7$ kJ mol$^{-1}$ [69].

*4.2 2-alkanol + n-alkanone, or + linear organic carbonate, or + linear ether systems*

*4.2.1 The effect of increasing the solvent size in mixtures with a given 2-alkanol*

At, this condition, $H_m^E$ decreases in systems containing a *n*-alkanone, or a carbonate, while $H_m^E$ increases in solutions with linear ethers (Table 2). For example, $H_m^E$ (2-propanol)/J mol$^{-1}$ = 1617 (propanone) [32]; 1504 (2-pentanone) [70]; 962 (dipropyl ether) [47]; 1065 (dibutyl ether) [71]. We note that $H_m^E$ values of 2-alkanol (fixed) + *n*-alkane mixtures increase with the alkane size, and that $H_m^E$ values of the systems *n*-alkanone, or linear organic carbonate, or linear ether + *n*-alkane (fixed) decrease when the size of the polar component increases along each homologous series. Some experimental values are now provided: $H_m^E$ (2-butanol)/J mol$^{-1}$ = 885 (hexane) [65], 1355 (hexadecane) [72] and $H_m^E$ (heptane)/J mol$^{-1}$ = 1676 (propanone) [73]; 1338 (2-butanone) [74]; 1135 (2-pentanone) [74]; 204 (DPE) [75]; 119 (DBE) [76]; 1988 (DMC) [77]; 1328 (DEC) [78]. These results show that $H_m^E$ values of mixtures formed by a given 2-alkanol and a linear ether or *n*-alkane change similarly and that the variation of the $H_m^E$ values of 2-alkanol (fixed) + *n*-alkanone, or + carbonate systems is the same to that of $H_m^E$ of the solutions *n*-alkanone, or + carbonate + *n*-alkane (fixed). That is, alkanol-alkanol interactions are determinant in systems with ethers, while solvent-solvent interactions are determinant in mixtures with *n*-alkanone or carbonate and a given 2-alkanol. In this case, the decrease of the $\Delta H_{\text{X-X}}$ term, consequence of the weakening of the dipolar interactions between solvent molecules when their size is increased (see, e.g. [42]), predominates over the increase of the $\Delta H_{\text{OH-OH}}$ and $\Delta H_{\text{OH-X}}$ contributions. The former is due to longer solvents are better breakers of the alkanol-alkanol interactions since they have larger aliphatic surfaces. The $\Delta H_{\text{OH-X}}$ contribution becomes less negative due to the X group is more sterically hindered in longer solvents and a lower number of interactions between unlike molecules are created upon mixing.

On the other hand, $H_m^E$ results of systems involving a given 2-alkanol change in the order: organic carbonate > $n$-alkanone > linear ether. This variation is the same that for mixtures formed by one of the mentioned polar compounds and heptane.

The application of the Flory model reveal that values of $\bar{\sigma}_r(H_m^E) = ((\sum \sigma_r(H_m^E))/N_S$; $N_S$, number of systems) change in the sequence: 0.280 (linear ether) > 0.086 (organic carbonate) > 0.063 ($n$-alkanone). That is, orientational effects become weaker in the mentioned order, and are much stronger in systems containing ethers, where the alcohol self-association may play a more relevant role. Note that the variation of the $X_{12}$ parameter with the composition is much sharper for the 2-propanol + DPE system than, e.g., in the solution with propanone at 363.15 K (Figure S1). It is interesting to show the values of $TS_m^E (= H_m^E - G_m^E)$ determined using the few experimental $G_m^E$ data available. Thus, $TS_m^E$/J mol$^{-1}$ = 955 (2-propanol + 2-butanone) [79], 1174 (2-butanol + 2-butanone) [80], 113 (2-propanol + DPE) [47], 345 (2-butanol + DPE) [31]. One can conclude that association effects are more important in solutions with ethers. The symmetry of the experimental $H_m^E(x_1)$ curves supports such statement since they are more shifted towards lower mole fractions of the alcohol ($x_1$) in the case of systems with linear monoethers, while the mentioned curves are more symmetrical for systems involving $n$-alkanones or linear organic carbonates (Figures 1-4). It is remarkable that the ERAS model has been applied, with rather good results, to 2-alkanol + DPE or + DBE systems [31,71,81]. We have performed ERAS calculations for $H_m^E$ of the 2-propanol + propanone mixture at 298.15 K using the following interaction parameters (determined in this work): $K_{AB} = 60$; $\Delta h_{AB}^* = -18.5$ kJ mol$^{-1}$; $\Delta v_{AB}^* = -8.6$ cm$^3$ mol$^{-1}$; $X_{AB} = 60$ J cm$^{-3}$. The theoretical $H_m^E(x_1)$ curve is very skewed towards low mole fractions of the alcohol (Figure S2), indicating that the model overestimates the alkanol self-association.

According to the Flory model, positive $X_{12}$ values indicate that interactions between like molecules are prevalent over those between unlike molecules. Inspection of Table 2 shows that this effect is strengthened in mixtures with linear organic carbonates and that becomes weaker in the order: linear organic carbonate > $n$-alkanone > linear monoether. Along 2-alkanol

(fixed) + *n*-alkanone or 2-butanol + linear organic carbonate systems, $X_{12}$ decreases when the solvent size increases. This suggests that the effect linked to the possible decrease of $\eta_{12}$ is overestimated by that related to the weakening of the dipolar interactions between solvent molecules ($\eta_{22}$ decrease). In contrast, the former effect is dominant in the systems 2-alkanol (fixed) + linear monoether, and $X_{12}$ increases with the chain length of the oxaalkane.

*4.2.2 The effect of replacing 2-propanol by 2-butanol in solutions with a given solvent*

At this condition, $H_m^E$ increases (Table 2), which be explained in similar terms that above: i.e., considering that the larger aliphatic surface of 2-butanol breaks more solvent-solvent interactions and that a lower number of interactions between unlike molecules is created along the mixing process when 2-butanol is involved since the OH group is more sterically hindered. Flory results show that $\sigma_r(H_m^E)$ values increase or remains nearly constant when 2-propanol is replaced by 2-butanol in systems with *n*-alkanones or DMC (Table 2). However, in solutions containing linear monoethers, $\sigma_r(H_m^E)$ decreases, which underlines that interactions between 2-propanol molecules are more relevant.

*4.3 Acetophenone systems*

In view of the available results on $H_m^E$ for 2-alkanol + *n*-alkanone mixtures (Table 2), one can expect that, when comparing $H_m^E$ values of solutions formed by a given 2-alkanol and acetophenone or *n*-alkanone of similar size, the results are higher for the mixtures including the aromatic ketone. In fact, this occurs for 1-alkanol + acetophenone mixtures [82]. Such behavior can be explained considering that the $\Delta H_{\text{CO-CO}}$ contribution is larger in solutions with acetophenone since its mixtures with *n*-alkane show miscibility gaps at temperatures close to 298.15 K. For example, the upper critical solution temperature (UCST) of the decane system is 277.4 K [82]. As a consequence, $H_m^E$(heptane)/J mol$^{-1}$ = 1493 (acetophenone) [83], 886 (2-heptanone) [84].

*4.4 Cyclization effect*

*4.4.1 Mixtures with cyclic ethers*

Firstly, we note that, in systems including a given 2-alkanol, $H_m^E$ values increase with the number of oxygen atoms of the ether (Table 2). Cyclic ether + heptane systems behave similarly. Thus, $H_m^E$(heptane)/ J mol$^{-1}$ = 815 (THF) [85], 1937 (1,3-dioxolane) [86], 607 (THP) [85], 1784 (1,4-dioxane) [87]. On the other hand, $H_m^E$(2-alkanol + cyclic ether) > $H_m^E$(cyclic ether + heptane). For example, $H_m^E$ (2-butanol)/J mol$^{-1}$ = 1147 (THF) [88], 2332 (1,3-

dioxolane) [89]. This suggests that 2-alkanols are good breakers of interactions between ether molecules. A similar trend is observed for 2-butanol + DMC mixtures (Table 2).

The application of the Flory model reveals that orientational effects are stronger in systems with THF or THP. Thus, $\bar{\sigma}_r(H_m^E)$ = 0.118 (cyclic monoethers), 0.049 (cyclic diethers). One can conclude that in the former mixtures, alkanol-alkanol interactions are more determinant, while in solutions with cyclic diethers interactions are mainly dispersive. Accordingly with this statement, $X_{12}$ remains roughly constant with the composition for the 2-propanol + 1,4-dioxane system (Figure S1). It is to be noted that the $H_m^E(x_1)$ curve of the 2-propanol + THF mixture is more skewed to lower mole fractions of the alcohol than for the corresponding solution with 1,4-dioxane (Figure 3)

*4.4.2 Mixtures with propylene carbonate*

These systems are also characterized by large and positive $H_m^E$ values. Note that interactions between propylene carbonate are rather strong in view of the large miscibility gaps of its systems with *n*-alkanes. Thus, at 313.1 K, the propylene carbonate + heptane mixture shows a miscibility gap for mole fractions of the carbonate between (0.001, 0.999) [90]. In contrast, the UCST of the dimethyl carbonate + decane system is 286.6 K [91]. This can explain the stronger orientational effects predicted by the Flory model for 2-alkanol + PC mixtures with regards to those present in systems with DMC or DEC (Table 2).

*4.5 Temperature dependence of $H_m^E$*

For the 2-propanol + heptane system, the $C_{pm}^E$ value is large and positive: 15.2 J mol$^{-1}$ K$^{-1}$ [92] remarking that the alcohol network is disrupted in large extent when the temperature is increased. This effect is strengthened when heptane is replaced by benzene or toluene. Thus, $(\frac{H_m^E}{\Delta T})_p$ / J mol$^{-1}$ K$^{-1}$ = 17.4 (benzene), 18 (toluene) [93]. That is, aromatic hydrocarbons are better breakers of the alkanol-alkanol interactions than alkanes. This is supported by the fact that for systems with a given 2-alkanol, say 2-propanol, $H_m^E$ (benzene) = 1284 [93] > $H_m^E$ (heptane) = 784 [94] (values in J mol$^{-1}$).

*4.5.1 Mixtures with ketones or linear organic carbonates*

According to experimental data, $(\frac{\Delta H_m^E}{\Delta T})_p$ (2-propanol)/J mol$^{-1}$ K$^{-1}$ = 3.7 (propanone) [95], 6.1 (2-butanone) [96], which indicates that association effects are rather weak in these solutions. On the other hand, $H_m^E$ of the 2-propanol + propanone mixture, increases in the temperature range (283.15-323.15) K and decreases at higher temperatures [95], i.e., dispersive interactions become dominant. Systems including acetophenone are characterized by larger $(\frac{\Delta H_m^E}{\Delta T})_p$ values (18.3 J mol$^{-1}$ K$^{-1}$ for the solution with 2-propanol [97]). This remarks that aromatic molecules are good breakers of the alcohol network, but also that dipolar interactions between acetophenone molecules are largely disrupted when the temperature is increased. This is supported by the fact that $H_m^E$ values of 2-alkanol + acetophenone mixtures are higher than the corresponding results for 2-alkanol, or acetophenone + hexane or + heptane systems. Thus, for the 2-butanol + acetophenone system, $H_m^E$ /J mol$^{-1}$ = 2260 [97], a value which is much higher than those of the mixtures acetophenone + heptane (1493 J mol$^{-1}$ [83]) or 2-butanol + hexane (885 J mol$^{-1}$ [65]). A similar trend is observed for systems with linear carbonates, since the $(\frac{\Delta H_m^E}{\Delta T})_p$ / J mol$^{-1}$ K$^{-1}$ values are also large and positive: 12.7 (2-propanol + DMC) [98]; 11.4 (2-butanol + DMC) [99]; 13.5 (2-butanol + DEC) [100].

*4.5.2 Mixtures with cyclic ethers*

In this case, the $(\frac{\Delta H_m^E}{\Delta T})_p$ results are lower or even negative. For systems with 2-butanol, such values, in J mol$^{-1}$ K$^{-1}$, are: −0.9 (THF) [88]; 4.0 (1,3-dioxolane); −0.2 (1,4-dioxane) [89]. It is clear that dispersive interactions are here more relatively important.

*4.5.3 Flory results at T > 298.15 K*

These results are collected, for a few systems, in Table S2. We note that $\sigma_r(H_m^E)$ values are lower than those obtained for the same mixtures at 298.15 K, showing the typical weakening of the orientational effects when $T$ is increased.

*4.6 Excess molar volumes and isochoric excess internal energies*

As already stated, most of the systems are characterized by positive $V_m^E$ values (Table S1). In addition, inspection of Tables 2 and S1 shows that $H_m^E$ and $V_m^E$ change in line along a given homologous series. Thus, it is possible to conclude that the main contribution to $V_m^E$ comes from interactional effects. Accordingly, $(\frac{\Delta V_m^E}{\Delta T})_p$ values are positive (Table S1). The Flory model provides much larger $V_m^E$ results than the experimental values (Table S1), indicating that the interactional contribution to this excess function is largely overestimated.

The isochoric molar excess internal energy, $U_{Vm}^E$ can be determined from the equation [48]:

$$U_{Vm}^E = H_m^E - \frac{T\alpha_p V_m^E}{\kappa_T} \tag{16}$$

As in previous applications, we have assumed along calculations that the mixtures show ideal behavior with regards to the expansion coefficient and isothermal compressibility. Table S1 lists the values of $U_{Vm}^{E}$ determined in this work. Since the $|V_{m}^{E}|$ values are usually small, the results of $U_{Vm}^{E}$ and $H_{m}^{E}$ are quite similar, and the relative variations of these magnitudes along a homologous series are also similar. For example, in the case of mixtures with 2-propanol, $H_{m}^{E}$/J mol$^{-1}$= 1114 (THF); 1119 (THP) [51] and $U_{Vm}^{E}$/J mol$^{-1}$ = 1027 (THF); 1072 (THP). It is interesting to note that the large $H_{m}^{E}$ value of the 2-butanol + DMC system (2743 J mol$^{-1}$ [99]) is in part due to structural effects since $U_{Vm}^{E}$ = 2474 J mol$^{-1}$.

### 4.7   Kirkwood-Buff integrals

We note that the curves $\delta_{11}^{0}$ vs $x_{1}$, the mole fraction of the alcohol, shows a large maximum for the 2-propanol + heptane mixture at 298.15 K (Figure 5), which can be ascribed to the alcohol self-association. When the alkane is replaced by a polar compound, the maximum value of $\delta_{11}^{0}$ largely decreases due to the new alkanol-solvent interactions created upon mixing Table 3, Figure 6). This shows that the alcohol network is strongly altered by the presence of a polar compound. For systems including 2-propanol, the mentioned maximum value of $\delta_{11}^{0}$ decreases in the order: linear monoether > propanone (or 2-butanone) > THF (Table 3, Figure 6). That is, effects related to alkanol-alkanol interactions are much more important in systems with linear monoethers. The symmetry of the $\delta_{11}^{0}(x_{1})$ curve of the 2-propanol + DPE mixture agrees with this statement. The low values of $|\delta_{ij}^{0}|$ for the solutions with THF, or $n$-alkanone indicate that, in the framework of this model, the behavior of these systems is close to random mixing, since they not form aggregates in solutions. Small preferential solvation effects have previously encountered in systems of alcohols with 1,4-dioxane [101], and somewhat stronger in mixtures with THF [102]. In the case of solutions with linear monoethers, our results suggest that their possible arrangement arises mainly from orientational effects rather that from association effects.

### 4.8   *Comparison with systems containing 1-propanol or 1-butanol*

Taking into account the limitations involved in the determination of the $\Delta H_{\text{OH-X}}^{\text{int}}$ values, results for mixtures containing 1-alkanols or 2-alkanols are very similar. Nevertheless, data suggest that interactions are stronger in mixtures with DMC. Inspection of Table 2 reveals that the variation of the $H_{m}^{E}$ values with the size of the solvent (characteristic group X), or of the 1-alkanol is similar to those previously discussed for systems with secondary alcohols, which show larger $H_{m}^{E}$ values. This is due to: (i) the $\Delta H_{\text{OH-X}}$ contribution is more negative for mixtures with 1-alkanols, since the OH group is less sterically hindered [2,58] and more interactions between unlike molecules are created upon mixing. (ii) The $\Delta H_{\text{OH-OH}}$ contribution is expected to be more positive in systems involving 2-alkanols. Note that for heptane mixtures, $H_m^E$/J mol$^{-1}$ = 614 (1-propanol) [103], 784 (2-propanol) [94]. The higher $H_m^E$ values of systems with 2-alkanols lead to the corresponding $V_m^E$ results are also higher. In addition, structural effects are more relevant in systems containing 1-alkanols since, for many systems, values of $H_m^E$ are > 0, and the corresponding $V_m^E$ results are < 0, or small in absolute value. For example, $V_m^E$(DPE)/cm$^3$ mol$^{-1}$ = $-0.391$ [104] (1-propanol), $-0.468$ (1-butanol) [105] (the corresponding results for 2-alkanol mixtures are given above); $V_m^E$(THP)/cm$^3$ mol$^{-1}$ = 0.0176 [106] (1-butanol); 0.282 [107] (2-butanol); $V_m^E$(2-butanone)/cm$^3$ mol$^{-1}$ = $-0.0298$ (1-propanol); 0.2197 (2-propanol) [79]; $V_m^E$(DEC)/cm$^3$ mol$^{-1}$ = 0.292 (1-butanol); 0.594 (2-butanol) [108]. On the other hand, there is a noticeable difference between the symmetry of the experimental $H_m^E(x_1)$ curves of systems with 1- or 2-alkanols and monoethers, since the former are more skewed to lower mole fraction of the alcohol $(x_1)$ (Figure 2). This means that association effects are more relevant in the mixtures with 1-propanol or 1-butanol. For the remainder solvents, the symmetry of such curves is quite similar (Figure S2). The values of $TS_m^E$ are consistent with this trend. Thus, $TS_m^E$/J mol$^{-1}$: $-74$ (1-propanol + DPE) [47]; $-14$ (1-butanol + DBE) [109,110]; 725 (1-propanol + 2-butanone) [79]; 812 (1-butanol + 2-butanone) [111]. Compare with the results given above for systems with 2-alkanols.

It is remarkable that, according to the results provided by the Flory model, no meaningful difference exists between orientational effects in systems with 1- or 2-alkanols and $n$-alkanone ($\bar{\sigma}_r(H_m^E)$(2-alkanol) = 0.063; $\bar{\sigma}_r(H_m^E)$(1-alkanol) = 0.051); or acetophenone ($\bar{\sigma}_r(H_m^E)$(2-alkanol) = 0.105; $\bar{\sigma}_r(H_m^E)$(1-alkanol) = 0.104); or cyclic ether ($\bar{\sigma}_r(H_m^E)$(2-alkanol) = 0.095; $\bar{\sigma}_r(H_m^E)$(1-alkanol) = 0.083), or linear organic carbonate ($\bar{\sigma}_r(H_m^E)$(2-alkanol) = 0.086; $\bar{\sigma}_r(H_m^E)$(1-alkanol) = 0.113). Larger differences exist for systems with propylene carbonate ($\bar{\sigma}_r(H_m^E)$(2-alkanol) = 0.172; $\bar{\sigma}_r(H_m^E)$(1-alkanol) = 0.139) and particularly for solutions with linear monoethers ($\bar{\sigma}_r(H_m^E)$ (2-alkanol) = 0.280; $\bar{\sigma}_r(H_m^E)$(1-alkanol = 0.375). That is, association effects seem to be are more relevant in 1-propanol or 1-butanol + DPE or + DBE systems than in the corresponding mixtures with 2-alkanols. In view of these results, it must be remarked that the dependence of the orienational effects with the functional group is the same along systems with 1- or 2-alkanols. For the solutions considered in Table 3, the formalism of the Kirkwood-Buff integrals does not provide meaningful differences between systems with 1-alkanols or 2-alkanols. This is not the case for 1-propanol or 2-propanol + heptane mixtures at 313.15 K, where the maximum values of $G_{11}$ ($x_1 \approx 0.1$)/cm$^3$ mol$^{-1}$ are approximately 11000 (1-propanol), and 5000 (2-propanol) [23].

Finally, we note that the values of the Flory interaction parameter are larger for mixtures with 2-alkanols, which can be interpreted as consequence of an increase of the 1-1 interactions and a decrease of the 1-2 interactions.

## 5.    Conclusions

Mixtures formed by 2-alkanol and an organic solvent characterized by a polar group (carbonyl, ether, or carbonate) have been studied using experimental data from the literature and by means of the application of the Flory model and the Kirkwood-Buff formalism. It has been shown that dipolar interactions between solvent molecules are determinant in solutions with *n*-alkanones or linear carbonate, and that alkanol-alkanol interactions are determinant in mixtures with linear monoethers. The application of the Flory model reveals that orientational effects in mixtures with a given 2-alkanol become weaker in the order: linear monoether > linear organic carbonate > *n*-alkanone, and that are stronger in systems with a cyclic monoether than in those with cyclic diethers or in solutions including acetophenone or propylene carbonate than in the corresponding mixtures involving linear solvents. These findings are consistent with results obtained using the Kirkwood-Buff integrals. According to the Flory model, orientational effects are similar in systems with 1- or 2-alkanols, except in the case of solutions with linear monoethers, where such effects are stronger when 1-alkanols are involved. No meaningful differences are encountered between mixtures with 1-alkanols or 2-alkanols when the formalism of the Kirkwood-Buff integrals is applied.

## 6.    References

[1]    C.G. Savini, D.R. Winterhalter, H.C. Van Ness. Heats of mixing of some alcohol-hydrocarbon systems. J. Chem. Eng. Data 10 (1968) 168-171.

[2]    I. Brown, W. Fock, F. Smith. The thermodynamic properties of solutions of normal and branched alcohols in benzene and *n*-hexane. J. Chem. Thermodyn. 1 (1969) 273-291.

[3]    E. Wolf, H. Wolff. Comparison of the association of isomeric propanols and propylamines. Fluid Phase Equilib. 17 (1984) 147-151.

[4]    M. Costas, D. Patterson. Self-association of alcohols in inert solvents. Apparent heat capacities and volumes of linear alcohols in hydrocarbons. J. Chem. Soc., Faraday Trans. I 81 (1985) 635-654.

[5]    M. Cáceres-Alonso, M. Costas, L. Andreolli-Ball, D. Patterson. Steric effects on the self-association of branched and cyclic alcohols in inert solvents. Apparent heat

capacities of secondary and tertiary alcohols in hydrocarbons. Can. J. Chem. 66 (1989) 989-998.

[6] G.M. Førland, F.O. Libnau, O.M. Kvalheim, H. Høiland. Self-association of medium-chain alcohols in *n*-decane solutions. Appl. Spectr. 50 (1996) 1264-1272.

[7] W.-C. Luo, J.-L. Lay, J.-S. Chen. NMR study of hydrogen bonding sssociation of some sterically hindered alcohols in carbon tetrachloride, chloroform and cyclohexane, Z. Phys. Chem. 216 (2002) 829-843.

[8] R. W. Larsen, M.A. Suhm. Cooperative organic hydrogen bonds: the librational modes of cyclic methanol clusters. J. Chem. Phys. 125 (2006) 154314-5.

[9] M.A. Czarnecki, M. Kwaśniewicz. Effect of the chain length on near infrared spectra of 1-alcohols from methanol to 1-decanol. NIR news. 30 (2019) 6-8.

[10] W.G. Killian, A.M. Bala, A.T. Norfleet, L. Peereboom, J.E. Jackson, C.T. Lira. Infrared quantification of ethanol and 1-butanol hydrogen bonded hydroxyl distributions in cyclohexane. Spectrochim. Acta, Part A: Mol. And Biomolecular Spectr. 285 (2023) 121837.

[11] G. Brink, L. Glasser. Dieletric studies of molecular association. A model for the association of ethanol in dilute solution. J. Phys. Chem. 82 (1978) 1000-1005.

[12] A. D'APrano, I.D. Donato, G. D'Arrigo, D. Bertolini, M. Cassettari, G. Salvetti. Molecular association and dynamics in *n*-pentanol and 2-methyl-2-butanol. Ultrasonic, dielectric and viscosity studies at different temperatures. Mol. Physics 55 (1985) 475-488.

[13] S. Schwerdtfeger, F. Köhler, R. Pottel, U. Kaatze. Dielectric relaxation of hydrogen bonded liquids: mixtures of monohydric alcohols with *n*-alkanes. J. Chem. Phys. 115 (2001) 4186-4194.

[14] M. El-Hefnawy, R. Tanaka. Density and relative permittivity for 1-alkanols + dodecane at 298.15 K. J. Chem. Eng. Data 50 (2005) 1651-1656.

[15] T.P. Iglesias, J.M. Forniés-Marquina, B. de Cominges. Excess permittivity of some mixtures *n*-alcohol + alkane: an interpretation of the underlying molecular mechanism. Mol. Phys. 103 (2005) 2639-2646.

[16] A. Nowok, M. Dulski, J. Grelska, A. Z. Szeremeta, K. Jurkiewicz, K. Grzybowska, M. Musiał, S. Pawlus. Phenyl ring: a steric hindrance or a source of different hydrogen bonding patterns in self-organizing systems? J. Phys. Chem. Lett. 12 (2021) 2142-2147.

[17] A. Heintz. A new theoretical approach for predicting excess properties of alkanol/alkane mixtures. Ber. Bunsenges. Phys. Chem. 89 (1985) 172-181.

[18] T. Vasiltsova, A. Heintz. New statistical mechanical model for calculating Kirkwood factors in self-associating liquid systems and its application to 1-alkanol + cyclohexane mixtures. J. Chem. Phys. 127 (2007) 114501.

[19]   J.A. González, I. García de la Fuente, J.C. Cobos, C. Casanova. A characterization of the aliphatic/hydroxyl interactions using a group contribution model (DISQUAC). Ber. Bunsenges. Phys. Chem. 95 (1991) 1658-1668.

[20]   J. Fernández, J.L. Legido, M.I. Paz Andrade, L. Pías, J. Ortega. Analysis of thermodynamic properties of 1-alkanol + *n*-alkane mixtures using the Nitta-Chao group contribution model. Fluid Phase Equilib. 55 (1990) 293-308.

[21]   W.A. Fouad, L. Wang, A. Haghmoradi, S.K. Gupta, W.G. Chapman. Understanding the Thermodynamics of hydrogen bonding in alcohol-containing mixtures: self-association. J. Phys. Chem. B 119 (2015) 14086-14101.

[22]   J.G. Kirkwood, F.P. Buff. The statistical theory of solutions. I. J. Chem. Phys. 19 (1954) 774-777.

[23]   J. Zielkiewicz. Kirkwood-Buff integrals in the binary and ternary mixtures containing heptane and aliphatic alcohol. J. Phys. Chem. 99 (1995) 3357-3364.

[24]   J.A. González, N. Riesco, I. Mozo, I. García de la Fuente, J.C. Cobos. Application of the Flory theory and of the Kirkwood-Buff formalism to the study of orientational effects in 1-alkanol + linear o cyclic monoether mixtures. Ind. Eng. Chem. Res., 48 (2009) 7417-7429.

[25]   J.A. González, A. Mediavilla, I. García de la Fuente, J.C. Cobos. Thermodynamics of (1-alkanol + linear polyether) mixtures. J. Chem. Thermodyn. 59 (2013) 195-208.

[26]   J.A. González, A. Mediavilla, I. García de la Fuente, J.C. Cobos, C. Alonso-Tristán, N. Riesco, Orientational effects in 1-alkanol + alkanone mixtures. Ind. Eng. Chem. Res. 52 (2013) 10317-10328.

[27]   J.A. González, F. Hevia, C. Alonso-Tristán, I. García de la Fuente, J.C. Cobos. Orientational effects in mixtures of organic carbonates with alkanes or 1-alkanols. Fluid Phase Equilib. 449 (2017) 91-103.

[28]   J.A. González, I. García de la Fuente, J.C. Cobos, C. Alonso-Tristán, L.F. Sanz. Orientational effects and random mixing in 1-alkanol + nitrile mixtures. Ind. Eng. Chem. Res. 54 (2015) 550-559.

[29]   P.J. Flory. Statistical Thermodynamics of liquid mixtures J. Am. Chem. Soc. 87 (1965) 1833-1838

[30]   A.B. Bhatia, D.E. Thornton. Structural aspects of the electrical resistivity of binary alloys. Phys. Rev. B2 (1970) 3004-3012.

[31]   R. Garriga, F. Sánchez, P. Pérez, M. Gracia. Vapour pressures at several temperatures of binary mixtures of di-*n*-propylether + 2-methyl-1-propanol, + 2-butanol or + 2-methyl-2-propanol, and excess functions at $T = 298.15$ K. Thermodynamic description according to the ERAS model. Ber. Bunsenges Phys. Chem. 102 (1994) 14-24.

[32]   I. Nagata. Excess enthalpies for (propan-2-ol + propanone) and (propan-2-ol + propanone + benzene) at the temperature 298.15 K. J. Chem. Thermodyn. 26 (1994) 691-695.

[33]   I. Nagata, K. Tamura, H. Kataoka, A. Ksiazczak. Excess enthalpies of ternary systems butan-1-ol or butan-2-ol + aniline + propanone at the temperature 298.15 K. J. Chem. Eng. Data 41 (1996) 593-597.

[34]   I. Nagata, A. Ksiazczak. Excess enthalpies of (ethanol or propan-2-ol + aniline + propanone) at the temperature 298.15 K. J. Chem. Thermodyn. 27 (1995) 1235-1240.

[35]   A. Abe, P.J. Flory. The thermodynamic properties of mixtures of small, nonpolar molecules. J. Am. Chem. Soc. 87 (1965) 1838-1846.

[36]   P.J. Flory, R.A. Orwoll, A. Vrij. Statistical thermodynamics of chain molecular liquids. I. An equation of state for normal paraffin hydrocarbons. J. Am. Chem. Soc. 86 (1964) 3507-3514.

[37]   A. Ben-Naim. Inversion of the Kirkwood-Buff theory of solutions: application to the water-ethanol system. J. Chem. Phys. 67 (1977) 4884-4890.

[38]   E. Matteoli, L. Lepori. Solute-solute interactions in water. II. An analysis through the Kirkwood-Buff integrals for 14 organic solutes. J. Chem. Phys. 80 (1984) 2856-2863.

[39]   Y. Marcus. Preferential solvation in mixed solvents. X. Completely miscible aqueous co-solvent binary mixtures at 298.15 K. Monatsh. Chem. 132 (2001) 1387-1411

[40]   J. Zielkiewicz. Solvation of DMF in the *N,N*-dimethylformamide + alcohol + water mixtures investigated by means of the Kirkwood-Buff integrals. J. Phys. Chem. 99 (1995) 4787-4793.

[41]   J. Zielkiewicz. Solvation of amide group by water and alcohols investigated using the Kirkwood-Buff theory of solutions. J. Chem. Soc., Faraday Trans. 94 (1998) 1713-1719.

[42]   F. Hevia, J.A. González, C. Alonso-Tristán, I. García de la Fuente, L.F. Sanz. Orientational effects in alkanone, alkanal or dialkyl carbonate + alkane mixtures, or in alkanone + alkanone, or + dialkyl carbonate systems. J. Mol. Liq. 233 (2017) 517-527.

[43]   J.A. González, I. Mozo, I. García de la Fuente, J.C. Cobos, V.A. Durov. Thermodynamics of 1-alkanol + cyclic ether mixtures. Fluid Phase Equilib. 245 (2006) 168-184.

[44]   P.J. Howell, B.J. Skillerne de Bristowe, D. Stubley. Enthalpies of mixing of carbon tetrachloride with some methyl-substituted benzenes. Part III. Analysis of the results by use of Flory's theory of liquid mixtures. J. Chem. Soc. A, (1971) 397-400.

[45]   J. Zielkiewicz. Preferential solvation of *N*-methylformamide, *N,N*-dimethylformamide and *N*-methylacetamide by water and alcohols in the binary and ternary mixtures. Phys. Chem. Chem. Phys. 2 (2000) 2925-2932.

[46] M. Almasi, H. Iloukhani. Densities, viscosities, and refractive indices of binary mixtures of acetophenone and 2-alkanols. J. Chem. Eng. Data 55 (2010) 1416-1420.

[47] R. Garriga, F. Sánchez, P. Pérez, M. Gracia. Isothermal vapour-liquid equilibria at eight temperatures and excess functions at 298.15 K of di-*n*-propylether with 1-propanol or 2-propanol. Fluid Phase Equilib. 138 (1997) 131-144.

[48] J.S. Rowlinson, F.L. Swinton. Liquids and Liquid Mixtures, Butterworths 3[th] Ed., London, 1982.

[49] H. Kalali, F. Kohler, P. Svejda. Excess properties of binary mixtures of 2,2,4-trimethylpentane with one polar component. Fluid Phase Equilib. 20 (1985) 75-80.

[50] H.P. Diogo, M.E. Minas de Piedade, J. Moura Ramos, J. Simoni, J.A. Martinho Simoes. Intermolecular forces in solution and lattice energies of ionic crystals. J. Chem. Educ. 70 (1993) A227-A233.

[51] T.M. Letcher, U.P. Govender. Excess molar enthalpies of an alkanol + a cyclic ether at 298.15 K. J. Chem. Eng. Data 40 (1995) 1097-1100.

[52] E. Calvo, P. Brocos, A. Piñeiro, M. Pintos, A. Amigo, R. Bravo, A.H. Roux, G. Roux.-Desgranges. Heat capacities, excess enthalpies, and volumes of binary mixtures containing cyclic ethers. 4. Binary systems 1,4-dioxane + 1-alkanols. J. Chem. Eng. Data 44 (1999) 948-954.

[53] J.A. González, I. Mozo, I. García de la Fuente, J.C. Cobos, N. Riesco. Thermodynamics of (1-alkanol + linear monoether) systems. J. Chem. Thermodyn. 40 (2008) 1495-1508.

[54] T.M. Letcher, B.C. Bricknell, B.C. Calorimetric investigation of the interactions of some hydrogen-bonded systems at 298.15 K. J. Chem. Eng. Data 41 (1996) 166-169.

[55] J.A. González, I. García de la Fuente, J.C. Cobos. Thermodynamics of mixtures containing oxaalkanes. 5. Ether + benzene, or + toluene systems. Fluid Phase Equilib. 301 (2011) 145-155.

[56] A. Liu, F. Kohler, L. Karrer, J. Gaube, P. Spelluci. A model for the excess properties of 1-alkanol + alkane mixtures containing chemical and physical terms. Pure & Appl. Chem. 61 (1989) 1441-1452.

[57] H. Renon, J.M. Prausnitz. On the thermodynamics of alcohol-hydrocarbon solutions. Chem. Eng. Sci. 22 (1967) 299-307.

[58] E.N. Rezanova, K. Kammerer, R.N. Lichtenthaler. Excess molar volumes and enthalpies of {an alkanol + *tert*-amyl methyl ether (TAME)}. J. Chem. Thermodyn. 32 (2000) 1569-1579.

[59] J.A. González, U. Domanska, J. Lachwa. Thermodynamics of binary mixtures containing a very strongly polar compound. 7. Isothermal VLE measurements for NMP + 2-propanol or + 2-butanol systems. DISQUAC and ERAS characterization of NMP or *N,N*-

dialkylamide + 2-alkanol mixtures. Comparison with results from Dortmund UNIFAC. Ind. Eng. Chem. Res. 44 (2005) 5795-5804.

[60] R.H. Stokes, C. Burfitt. Enthalpies of dilution and transfer of ethanol in non-polar solvents. J. Chem. Thermodyn. 5 (1973) 623-631.

[61] S.J. O'Shea, R.H. Stokes. Activity coefficients and excess partial molar enthalpies for (ethanol + hexane) from 283 to 318 K. J. Chem. Thermodyn. 18 (1986) 691-696.

[62] H.C. Van Ness, J. Van Winkle, H.H. Richtol, H.B. Hollinger. Infrared spectra and the thermodynamics of alcohol-hydrocarbon systems. J. Phys. Chem. 71 (1967) 1483-1494.

[63] M. Woycicka, B. Kalinowska. Bull Acad. Pol, Sci., Ser. Sci. Chim. 25 (1977) 639-647.

[64] D.M. Trampe, C.A. Eckert. Calorimetric measurement of partial molar excess enthalpies at infinite dilution. J. Chem. Eng. Data 36 (1991) 112-118.

[65] S. Murakami, K. Amaya, R. Fujishiro. Heats of mixing for binary mixtures. The energy of hydrogen bonding between alcohol and ketone molecules. Bull. Chem. Soc. Jpn. 37 (1964) 1776-1780.

[66] I. Nagata. Excess enthalpies of (aniline + butan-1-ol) and of (aniline + butan-1-ol + benzene) at the temperature 298.15 K. J. Chem. Thermodyn. 25 (1993) 1281-1285.

[67] M.M. H. Bhuiyan, K. Tamura. Excess molar enthalpies of ternary mixtures of (methanol, ethanol + 2-propanol + 1,4-dioxane) at $T$ = 298.15 K. J. Chem. Thermodyn. 36 (2004) 549-554.

[68] F. Takahashi, Y. Sakai, Y. Nakazawa, Y. Mizutami. The change of diamagnetic susceptibility and the thermodynamic data due to hydrogen bonding between 2-propanol and various hydrogen acceptors. Bull Chem. Soc. Jpn. 67 (1994) 2967-2971.

[69] K.R. Patil, G. Pathak, SD. Pradhan. Excess enthalpies and excess volumes of mixing for isomeric butanol-acetone systems at 303.15 K. Thermochim. Acta 177 (1991) 143-149.

[70] T.M. Letcher, J.A. Nevines. Excess enthalpies of ketone + alkanol at the temperature 298.15 K. J. Chem. Eng. Data 40 (1995) 995-996.

[71] E.N. Rezanova, K. Kammerer, R.N. Lichtenthaler. Excess properties of binary alkanol + diisopropylether (DIPE) or + dibutyl ether (DBE) mixtures and the application of the Extended Real Associated Solution model. J. Chem. Eng. Data 44 (1999) 1235-1239.

[72] J. Jiménez, J. Valero, M. Gracia, C. Gutiérrez Losa, Excess molar enthalpy of (an *n*-alkane + a butanol isomer). J. Chem. Thermodyn. 20 (1988) 931-936.

[73] Y. Akamatsu, H. Ogawa, S. Murakami, Molar excess enthalpies, molar excess volumes and molar isentropic compressions of mixtures of 2-propanone with heptane, benzene and trichloromethane at 298.15 K, Thermochim. Acta, 113 (1987) 141-150.

[74]  O. Kiyohara, Y.P. Handa, G.C. Benson, Thermodynamic properties of binary mixtures containing ketones III. Excess enthalpies of *n*-alkanes + some aliphatic ketones. J. Chem. Thermodyn. 11 (1979) 453-460.

[75]  F. Kimura, P.J. D'Arcy, G.C. Benson. Excess enthalpies and the heat capacities for (di-*n*-propylether + *n*-heptane). J. Chem. Thermodyn. 15 (1983) 511-516.

[76]  G.C. Benson, B. Luo, B.C.-Y. Lu. Excess enthalpies of dibutyl ether + *n*-alkane at 298.15 K. Can. J. Chem. 66 (1988) 531–534.

[77]  I. García, J.C. Cobos, J.A. González, C. Casanova, M.J. Cocero, Thermodynamics of binary mixtures containing organic carbonates. 1. Excess enthalpies of dimethyl carbonate + hydrocarbons or + tetrachloromethane. J. Chem. Eng. Data 33 (1988) 423-426.

[78]  I. García, J.C. Cobos, J.A. González, C. Casanova, Excess enthalpies of diethyl carbonate + some normal alkanes ($C_6$ - $C_{14}$) + cyclohexane, + methylcyclohexane, + benzene, + toluene, or + tetrachloromethane, Int. DATA Ser., Sel. Data Mixtures, Ser. A, 15 (1987) 164-173.

[79]  R. Garriga, F. Sánchez, P. Pérez, M. Gracia. Excess Gibbs free energies at seven temperatures and excess enthalpies and volumes at 298.15 K of butanone with 1-propanol, or 2-propanol. Fluid Phase Equilib. 124 (1996) 123-134.

[80]  R. Garriga, S. Martínez, P. Pérez, M. Gracia. Vapour pressures at several temperatures, and excess functions at $T$ = 298.15 K of (butanone + 2-butanol). J. Chem. Thermodyn. 31 (1999) 117-127.

[81]  K. Kammerer, R.N. Lichtenthaler. Excess properties of binary alkanol-ether mixtures and the application of the ERAS model. Thermodchim. Acta 310 (1998) 61-67.

[82]  J.A. González, C. Alonso-Tristán, I. García de la Fuente, J.C. Cobos. Liquid-liquid equilibria for acetophenone + *n*-alkane mixtures and characterization of acetophenone system using DISQUAC. Fluid Phase Equilib. 391 (2015) 39-48.

[83]  O. Urdaneta, S. Haman, Y.P. Handa, G.C. Benson. Thermodynamic properties of binary mixtures containing ketones. IV. Excess enthalpies of acetophenone + an *n*-alkane and phenylacetone + *n*-alkane. J. Chem. Thermodyn. 11 (1979) 851-856.

[84]  O. Urdaneta, Y.P. Handa, G.C. Benson, Thermodynamic properties of binary mixtures containing ketones V. Excess enthalpies of an isomeric heptanone + *n*-heptane, J. Chem. Thermodyn. 11 (1979) 857-860.

[85]  I. Castro, M. Pintos, A. Amigo, R. Bravo, M.I. Paz Andrade. Excess enthalpies of (tetrahydrofuran or tetrahydropyran + an *n*-alkane) at the temperature 298.15 K. J. Chem. Thermodyn. 26 (1994) 29-33.

[86] P. Brocos, E. Calvo, A. Amigo, R. Bravo, M. Pintos, A. Roux, G. Roux-Desgranges. Heat capacities, excess enthalpies, and volumes of mixtures containing cyclic ethers. 2. Binary systems 1,3-dioxolane + *n*-alkanes. J. Chem. Eng. Data 43 (1998) 112-116.

[87] E. Calvo, P. Brocos, R. Bravo, M. Pintos, A. Amigo A. Roux, G. Roux-Desgranges. Heat capacities, excess enthalpies, and volumes of mixtures containing cyclic ethers. 2. Binary systems 1,4-dioxane + *n*-alkanes J. Chem. Eng. Data 43 (1998) 105-111.

[88] A. Valen, M.C. López, J.S. Urieta, F.M. Royo, C. Lafuente. Thermodynamic study of mixtures containing oxygenated compounds. J. Mol. Liq. 95 (2002) 157-165.

[89] I. Gascón, H. Artigas, S. Martín, P. Cea, C. Lafuente. Excess molar enthalpies of 1,3-dioxolane, or 1,4-dioxane with isomeric butanols. J. Chem. Thermodyn. 34 (2002) 1351-1360.

[90] J. Li, Q. Zhao, X. Tang, K. Xiao, J. Yuan. Liquid-liquid equilibria for the systems: heptane + benzene + solvent (propylene carbonate, *N,N*-dimethylformamide, or mixtures ) at temperatures from (303.2 to 323.2) K. J. Chem. Eng. Data 59 (2014) 3307-3313.

[91] J.A. González, I. García, J.C. Cobos, C. Casanova. Thermodynamics of binary mixtures containing organic carbonates. 4. Liquid-liquid equilibria of dimethyl carbonate + selected *n*-alkanes. J. Chem. Eng. Data 36 (1991) 162-164.

[92] R. Tanaka, S. Toyama. Excess molar heat capacities and excess molar volumes of (propan-2-ol, or butan-2-ol, or pentan-2-ol, or penta-3-ol, or 2-methylbutan-2-ol + *n*-heptane) at the temperature 298.15 K. J. Chem. Thermodyn. 28 (1996) 1403-1410.

[93] H.C. Van Ness and M.M. Abbott. Int. DATA Ser. Sel. Data Mixtures Ser. A, 1, (1976).

[94] A. Heintz. The pressure dependence of the excess enthalpy of mixtures of isopropanol with heptane and isooctane up to 553 bar. Ber. Bunsenges. Phys. Chem. 85 (1981) 632-635.

[95] B. Lowen, S. Schulz. Excess molar enthalpies of acetone + water, cyclohexane, methanol, 1-propanol, 2-propanol, 1-butanol and 1-pentanol at 283.15, 298.15, 323.15, 343.15 and 363.15 K. Thermochim. Acta 262 (1995) 69-82.

[96] I. Nagata, T. Ohta, S. Nakagawa. Excess Gibbs free energy and heats of mixing for binary liquid mixtures of alcohols. J. Chem. Eng. Jpn. 9 (1976) 276-281.

[97] S. Li, W. Yan. Excess molar enthalpies of acetophenone + (methanol, ethanol, + 1-propanol, and + 2-propanol) at different temperatures and pressures. J. Chem. Eng. Data 53 (2008) 551-555.

[98] S. Li, H. Dong, W. Yan, B. Peng. Excess molar enthalpies of dimethyl carbonate and (methanol, ethanol, + 1-propanol, and + 2-propanol) at $T$ = 298.15, 313.15 and 328.15) K and $p$ = (0.1, 1.0 and 10.0) MPa. J. Chem. Eng. Data 50 (2005) 1087-1090.

[99] R. Francesconi, F. Comelli. Excess molar enthalpies and excess molar volumes of binary mixtures containing dimethyl carbonate + four butanol isomers at (288.15, 298.15, and 313.15) K. J. Chem. Eng. Data 44 (1999) 44-47.

[100] F. Comelli, R. Francesconi, G. Castellari. Excess molar enthalpies of diethyl carbonate + four butanol isomers in the range (288.15-318.15) K. J. Chem. Eng. Data 44 (1999) 739-743.

[101] Y. Marcus. Preferential solvation in mixed solvents. 14. Mixtures of 1,4-dioxane with organic solvents: Kirkwood–Buff integrals and volume-corrected preferential solvation parameters. J. Mol. Liq. 128 (2006) 115-126.

[102] Y. Marcus. Preferential solvation in mixed solvents 13. Mixtures of tetrahydrofuran with organic solvents: Kirkwood-Buff integrals and volume-corrected preferential solvation parameters. J. Solution Chem. 35 (2006) 251-277.

[103] F.E.M. Alaoui, F. Aguilar, M.J. González-Fernández, A. El Amarti, E. Montero Oxygenated compounds + hydrocarbon mixtures in fuels and biofuels: excess enthalpies of ternary mixtures containing 1-butoxybutane + propan-1-ol + hex-1-ene, or heptane, or 2,2,4-trimethylpentane at (293.15 and 313.15) K. J. Chem. Eng. Data 59 (2014) 2856-2864.

[104] A. Serna, I. García de la Fuente, J.A. González, J.C. Cobos, C. Casanova. Excess molar volumes of 1-alcohol + aliphatic monoethers at 298.15 K. Fluid Phase Equilib. 110 (1995) 361-367.

[105] I. Iñarrea, J. Valero, P. Pérez, M. Gracia, C. Gutiérrez Losa. $H_m^E$ and $V_m^E$ of some (butanone or dipropylether + an alkanol) mixtures. J. Chem. Thermodyn. 20 (1988) 193-199.

[106] Y. Miranda, A. Piñeiro, P. Brocos. Thermodynamics of mixing tetrahydropyran with 1-alkanols and excess enthalpies of homomorphy-related systems. J. Chem. Eng. Data 52 (2007) 429-437.

[107] H. Quinteros-Lama, M. Cartes, A. Mejía, H. Segura. Experimental determination and theoretical modeling of the vapor-liquid equilibrium and densities of the binary system bultan-2-ol + tetrahydro-H-pyran. Fluid phase Equilib. 342 (2013) 52-59.

[108] A. Rodríguez, J. Canosa, J. Tojo. Density, refractive index, and speed of sound of binary mixtures (diethyl carbonate + alcohols) at several temperatures J. Chem. Eng. Data 46 (2001) 1506-1515.

[109] M.A. Villamañán, C. Casanova, A.H. Roux, J.-P.E. Grolier. Calorimetric investigation of the interactions between oxygen and hydroxyl groups in (alcohol + ether) at 298.15 K. J. Chem. Thermodyn., 14 (1982) 251-258.

[110] L. Bernazzi, M.R. Carosi, N. Ceccanti, G. Conti, P. Gianni, V. Mollica, M.R. Tiné, L. Lepori, E. Matteoli. Thermodynamic study of organic compounds in di-*n*-butylether. Enthalpy and Gibbs energy of solvation. Phys. Chem. Chem. Phys. 2 (2000) 4829-4836.

[111] R. Garriga, F. Sánchez, P. Pérez, M. Gracia. Isothermal vapor-liquid equilibria of butanone + butan-1-ol between 278.15 and 323.15 K. J. Chem. Eng. Data 41 (1996) 451-454.

[112] B. González, A. Domínguez, J. Tojo. Viscosities, densities and sepeeds of sound of the binary systems: 2-propanol with octane, or decane, or dodecane at $T$ = (293.15, 298.15 and 303.15) K. J. Chem. Thermodyn. 35 (2003) 939-953.

[113] G. Roux, D. Roberts, G. Perron, J.E. Desnoyers. Microheterogeneity in aqueous-organic solutions: heat capacities, volumes and expansibilities of some alcohols, aminoalcohol and tertiary amines in water, J. Solution Chem. 9 (1980) 629-647.

[114] B. González, A. Domínguez, J. Tojo. Dynamic viscosities of 2-butanol with alkanes ($C_8$, $C_{10}$, and $C_{12}$) at several temperatures. J. Chem. Thermodyn. 36 (2004) 267-275.

[115] T. Okano, H. Ogawa, S. Murakami. Molar excess volumes, isentropic compressions, and isobaric heat capacities of methanol-isomeric butanol systems at 298.15 K. Can. J. Chem. 66 (1988) 713-717.

[116] M.N. Roy, B.K. Sarkar, R. Chanda. Viscosity, density, and speed of sound for the binary mixtures of formamide with 2-methoxyetahnol, acetophenone, acetonitrile,1,2-dimethoxyethane and dimethylsulfoxide at different temperatures. J. Chem. Eng. Data 52 (2007) 1630-1637.

[117] M.M. Phillip. Adiabatic and isothermal compressibilities of liquids. Proc. Indian Acad. Sci. A9 (1939) 109-120.

[118] B.A. Coomber, C.J. Wormald. A stirred flow calorimeter. The excess enthalpies of acetone + water and of acetone + some normal alcohols. J. Chem. Thermodyn. 8 (1976) 793-799.

[119] M. López, J.L. Legido, L. Romaní, E. Carballo, E. Pérez Martell, E. Jímenez, M.I. Paz Andrade. Excess molar enthalpies for (propan-1-ol + pentan-3-one + hexane) at the temperature 298.15 K. J. Chem. Thermodyn. 24 (1992) 205-212.

[120] J.P. Chao, M. Dai, M. Studies of thermodynamic properties of binary mixtures containing an alcohol. XVI. Excess molar enthalpies of each of (one of the four butanols + methyl ethyl ketone or methyl isobutyl ketone) at the temperature 298.15 K. J. Chem. Thermodyn. 23 (1991) 117-121.

[121] S. Li, H. Gao, W. Yan. Determination and correlation of excess moalr enthalpies of eight binary systems containing acetophenone at different temperatures. J. Chem. Eng. Data, 53 (2008) 1630-1634.

[122]  I. Mozo, I. García de la Fuente, J.A. González, J.C, Cobos. Molar excess enthalpies at 298.15 K for 1-alkanol + dibutylether systems. J. Chem. Thermodyn. 42 (2010) 17-22.

[123]  P. Brocos, E. Calvo, A. Piñeiro, R. Bravo, A. Amigo, A.H. Roux, G. Roux-Desgranges. Heat capacities, excess enthalpies, and volumes of mixtures containing cyclic ethers. 5. Binary systems {1,3-dioxolane + 1-alkanols}. J. Chem. Eng. Data 44 (1999) 1341-1347.

[124]  C. Vallés, E. Pérez, M. Cardoso, M. Domínguez, A.N. Mainar. Excess enthalpy, density, viscosity, and speed of sound for the mixture tetrahydropyran + 1-butanol at (283.15, 298.15, and 313.15) K. J. Chem. Eng. Data 49 (2004) 1460-1464.

[125]  R. Francesconi, F. Comelli. Excess molar enthalpies, densities, and excess molar volumes of diethyl carbonate in binary mixtures with seven n-alkanols at 298.15 K. J. Chem. Eng. Data 42 (1997) 45-48.

[126]  R. Francesconi, F. Comelli. Excess enthalpies and excess volumes of the liquid binary mixtures of propylene carbonate + six alkanols at 298.15 K. J. Chem. Eng. Data 41 (1996) 1397-1400.

[127]  J.A. Salas, E.L Arancibia, M. Katz. Excess molar volumes and isothermal vapor-liquid equilibria in the tetrahydrofuran with propan-1-ol and propan-2-ol systems at 298.15 K. Can. J. Chem. 75 (1997) 207-211.

[128]  R. Garriga, P. Pérez, M. Gracia. Vapour pressures at eight temperatures of mixtures of di-*n*-propylether + ethanol or + 1-butanol. Thermodynamic description of mixtures of di-*n*-propylether + alkanol according to the ERAS model. *Ber Bunsenges Phys. Chem.* 101 (1997) 1466-1473.

[129]  S. Martínez, R. Garriga, P. Pérez, M Gracia. Densities and viscosities of binary mixtures of butanone with butanol isomers at several temperatures. Fluid Phase Equilib. 168 (2000) 267-279.

[130]  M. Keller, S. Schnabel, A. Heintz. Thermodynamics of the ternary mixture propan-1-ol + tetrahydrofuran + *n*-heptane at 298.15 K. Experimental results and ERAS model calculations of $G^E$, $H^E$ and $V^E$. Fluid Phase Equilib. 110 (1995) 231-265.

[131]  A. Heintz, E. Dolch, R.N. Lichtenthaler. New experimental VLE data for alkanol/alkane mixtures and their description by a extended real association (ERAS) model. Fluid Phase Equilib. 27 (1986) 61-79.

**TABLE 1**

Physical constants and Flory parameters[a] of pure compounds at $T$ = 298.15 K.

| compound | $V_{mi}$/cm$^3$ mol$^{-1}$ | $\alpha_{pi}$/$10^{-3}$ K$^{-1}$ | $\kappa_{Ti}$/$10^{-12}$ Pa$^{-1}$ | $V_{mi}^*$/cm$^3$ mol$^{-1}$ | $P_i^*$/J cm$^{-3}$ |
|---|---|---|---|---|---|
| 2-propanol | 76.96 [112] | 1.09 [112] | 1147 [112,113] | 60.80 | 454 |
| 2-butanol | 92.34 [115] | 1.03 [114] | 995 [114,115] | 73.65 | 485.2 |
| acetophenone | 117.44 [117] | 0.81 [116] | 682 [116,117] | 97.27 | 515.8 |

[a]$V_{mi}$, molar volumen; $\alpha_{pi}$, isobaric expansion coefficient; $\kappa_{Ti}$, isothermal compressibility, $V_{mi}^*$, $P_i^*$, reduction parameter for volume and pressure

**TABLE 2**

Excess molar enthalpies, $H_m^E$, at equimolar composition and 298.15 K, for alkanol (1) + organic solvent (2) mixtures. Values of the Flory interaction parameter, $X_{12}$, and of the relative standard deviations for $H_m^E$, $\sigma_r(H_m^E)$ (equation 13) are also included.

| System[a] / $X_{12}$ /J cm$^{-3}$ | $N^b$ | $H_m^E$/ J mol$^{-1}$ | $\sigma_r(H_m^E)$ | Ref | System[a] / $X_{12}$ /J cm$^{-3}$ | $N^b$ | $H_m^E$/ J mol$^{-1}$ | $\sigma_r(H_m^E)$ | Ref |
|---|---|---|---|---|---|---|---|---|---|
| 2-alkanol + ketone | | | | | 1-alkanol + ketone | | | | |
| 2PrOH + 1CO1 103.86 | 19 | 1617 | 0.082 | 32 | 1PrOH + 1CO1 90.12 | 19 | 1365 | 0.038 | 118 |
| 2PrOH + 1CO2 88.67 | 12 | 1482 | 0.027 | 79 | 1PrOH + 1CO2 75.21 | 19 | 1237 | 0.034 | 79 |
| 2PrOH + 1CO3 84.95 | 10 | 1504 | 0.037 | 70 | 1PrOH + 1CO3 69.22 | 19 | 1209 | 0.085 | 70 |
| 2PrOH + 2CO2 83.00 | 10 | 1465 | 0.053 | 70 | 1PrOH + 2CO2 70.74 | 19 | 1232 | 0.101 | 119 |
| 2BuOH + 1CO1 106.67 | 17 | 1865 | 0.108 | 33 | 1BuOH +1CO1 89.49 | 19 | 1537 | 0.015 | 118 |
| 2BuOH + 1CO2 86.34 | 13 | 1630 | 0.071 | 80 | 1BuOH +1CO2 75.55 | 19 | 1421 | 0.033 | 120 |
| 2PrOH + ACP 86.93 | 12 | 1568 | 0.100 | 97 | 1PrOH + ACP 75.85 | 12 | 1364 | 0.109 | 97 |
| 2BuOH + ACP 108.91 | 14 | 2260 | 0.111 | 97 | 1BuOH + ACP 68.80 | 13 | 1434 | 0.099 | 121 |
| 2-alkanol + ether | | | | | 1-alkanol + ether | | | | |
| 2PrOH + DPE 50.84 | 14 | 962 | 0.274 | 47 | 1PrOH + DPE 39.85 | 13 | 740 | 0.377 | 47 |
| 2PrOH + DBE 53.07 | 9 | 1065 | 0.393 | 71 | 1PrOH + DBE 44.71 | 21 | 886 | 0.414 | 122 |
| 2BuOH + DPE 49.58 | 15 | 1069 | 0.219 | 31 | 1BuOH + DPE 34.96 | 12 | 742 | 0.346 | 105 |
| 2BuOH + DBE 52.68 | 9 | 1215 | 0.235 | 81 | 1BuOH + DBE 37.57 | 27 | 865 | 0.363 | 122 |
| 2PrOH + THF 68.75 | 13 | 1114 | 0.154 | 51 | 1PrOH + THF 58.40 | 18 | 933 | 0.084 | 51 |

TABLE 2 (continued)

| | | | | | | | | | |
|---|---|---|---|---|---|---|---|---|---|
| 2PrOH + THP 65.24 | 18 | 1119 | 0.170 | 51 | 1PrOH + THP 54.23 | 17 | 922 | 0.185 | 51 |
| 2PrOH + 14DX 124.89 | 19 | 2066 | 0.055 | 51 | 1PrOH + 14DX 108.10 | 17 | 1769 | 0.037 | 52 |
| 2PrOH + 13DX | | | | | 1PrOH + 13DX 121.56 | 17 | 1858 | 0.039 | 123 |
| 2BuOH + THF 62.79 | 11 | 1147 | 0.099 | 58 | 1BuOH + THF 51.04 | 11 | 927 | 0.092 | 88 |
| 2BuOH + THP | | | | | 1BuOH + THP 46.79 | 11 | 923 | 0.157 | 124 |
| 2BuOH + 14DX 115.16 | 11 | 2172 | 0.057 | 89 | 1BuOH +14DX 105.21 | 17 | 1975 | 0.030 | 52 |
| 2BuOH + 13DX 133.95 | 11 | 2335 | 0.035 | 89 | 1BuOH +13DX 115.34 | 17 | 2011 | 0.038 | 123 |
| | 2-alkanol + organic carbonate | | | | 1-alkanol + organic carbonate | | | | |
| 2PrOH + DMC 133.49 | 17 | 2195 | 0.069 | 98 | 1PrOH + DMC 120.34 | 19 | 1955 | 0.073 | 98 |
| 2PrOH + DEC | | | | | 1PrOH + DEC 99.09 | 19 | 1794 | 0.166 | 125 |
| 2BuOH + DMC 147.30 | 17 | 2743 | 0.096 | 99 | 1BuOH + DMC 126.63 | 19 | 2356 | 0.082 | 99 |
| 2BuOH + DEC 109.72 | 17 | 2291 | 0.093 | 100 | 1BuOH + DEC 93.19 | 19 | 1944 | 0.134 | 125 |
| 2PrOH + PC 144.21 | 18 | 2375 | 0.149 | 126 | 1PrOH + PC 130.69 | 19 | 2138 | 0.126 | 126 |
| 2BuOH + PC 128.30 | 17 | 2405 | 0.195 | 126 | 1BuOH + PC 116.66 | 19 | 2200 | 0.153 | 126 |

[a]symbols are: 2PrOH, 2-propanol; 2BuOH, 2-butanol; 1PrOH, 1-propanol; 1BuOH, 1-butanol; 1CO1, propanone; 1CO2, 2-butanone; 1CO3, 2-pentanone; 2CO2, 3-pentanone; ACP, acetophenone; DPE, dipropyl ether; DBE, dibutyl ether; THF, oxolane; THP, oxane; 14DX. 1,4-dioxane; 13DX, 1,3-dioxolane; DMC, dimethyl carbonate; DEC, diethyl carbonate; PC, propylene carbonate; PC; [b]number of data points

**TABLE 3**

Linear coefficients of preferential solvation, $\delta_{ij}^0$, at 2981.5 K and composition $x_1$ for alkanol (1) + organic solvent (2) mixtures.

| System[a] | $x_1$ | $\delta_{11}^0$ /cm$^3$ mol$^{-1}$ | $\delta_{12}^0$ /cm$^3$ mol$^{-1}$ | Ref. |
|---|---|---|---|---|
| 2-PrOH(1) + DPE(2) | 0.2 | 169.3 | −12.6 | 47 |
| | 0.4 | 173.9 | −44.7 | |
| | 0.6 | 83.7 | −43.2 | |
| | 0.8 | 27.1 | −29.2 | |
| 2-BuOH(1) + DPE(2) | 0.2 | 123.2 | −12.1 | 31 |
| | 0.4 | 124.4 | −39.8 | |
| | 0.6 | 60.6 | −39.0 | |
| | 0.8 | 19.2 | −25.0 | |
| 2-PrOH(1) + 1CO2(2) | 0.2 | 27.6 | −3.5 | 79 |
| | 0.4 | 36.5 | −15.9 | |
| | 0.6 | 27.9 | −28.4 | |
| | 0.8 | 10.4 | −25.7 | |
| 2-BuOH(1) + 1CO2(2) | 0.2 | 18.8 | −5.5 | 80 |
| | 0.4 | 31.8 | −22.8 | |
| | 0.6 | 21.3 | −33.9 | |
| | 0.8 | 4.7 | −20.3 | |
| 2-PrOH(1) + THF(2) | 0.2 | 13.2 | −2.3 | 127 |
| | 0.4 | 13.7 | −6.8 | |
| | 0.6 | 7.7 | −8.0 | |
| | 0.8 | 2.4 | −4.9 | |
| 1-PrOH(1) + DPE(2) | 0.2 | 159.0 | −9.9 | 47,104 |
| | 0.4 | 164.6 | −39.0 | |
| | 0.6 | 79.5 | −37.9 | |
| | 0.8 | 25.7 | −24.9 | |
| 1-BuOH(1) + DPE(2) | 0.2 | 149.9 | −16.1 | 128,105 |
| | 0.4 | 129.5 | −41.6 | |
| | 0.6 | 57.1 | −35.4 | |
| | 0.8 | 18.8 | −24.2 | |

TABLE 3 (continued)

| | | | | |
|---|---|---|---|---|
| 1PrOH(1) + 1CO2(2) | 0.2 | 30.7 | −3.5 | 79 |
| | 0.4 | 35.1 | −13.9 | |
| | 0.6 | 25.2 | −23.4 | |
| | 0.8 | 9.9 | −22.8 | |
| 1BuOH(1) + 1CO2(2) | 0.2 | 24.5 | −6.6 | 111,129 |
| | 0.4 | 33.9 | −23.8 | |
| | 0.6 | 23.4 | −36.8 | |
| | 0.8 | 6.5 | −28.0 | |
| 1PrOH(1) + THF(2) | 0.2 | 16.2 | −3.1 | 130 |
| | 0.4 | 19.4 | −9.3 | |
| | 0.6 | 10.0 | −10.0 | |
| | 0.8 | 2.8 | −5.3 | |

[a]for symbols, see Table 2.

**TABLE 4**

Partial molar excess enthalpies at infinite dilution, $H_1^{E,\infty}$, at $T = 298.15$ K at atmospheric pressure for solute(1) + organic solvent(2) mixtures, and hydrogen bond enthalpies, $\Delta H_{OH-X}^{int}$, for alkanol (1) + organic solvent (2) systems

| System[a] | $H_{m1}^{E,\infty}$ / kJ mol$^{-1}$ | $\Delta H_{OH-X}^{int}$ / kJ mol$^{-1}$ | System[a] | $H_{m1}^{E,\infty}$ / kJ mol$^{-1}$ | $\Delta H_{OH-X}^{int}$ / kJ mol$^{-1}$ |
|---|---|---|---|---|---|
| 1CO1 + heptane | 9.1 [73] | | | | |
| 1CO2 + heptane | 7.5 [74] | | | | |
| 1CO3 + heptane | 6.4 [74] | | | | |
| 2CO2 + heptane | 5.9 [74] | | | | |
| ACP + heptane | 9.2 [83] | | | | |
| DPE + heptane | 0.84 [75] | | | | |
| DBE + heptane | 0.53 [76] | | | | |
| THF + heptane | 3.2 [85] | | | | |
| THP + heptane | 2.2 [85] | | | | |
| 1,4-DX + heptane | 7.3 [87] | | | | |
| 1,3-DX + heptane | 7.6 [86] | | | | |
| DMC + heptane | 9.5 [77] | | | | |
| DEC + heptane | 7.1 [78] | | | | |
| 2PrOH + 1CO1 | 7.1 [32] | −25.2 | 1PrOH + 1CO1 | | −26.0 [26] |
| 2PrOH + 1CO2 | 6.0 [79] | −24.7 | 1PrOH + 1CO2 | | −25.4 [26] |
| 2PrOH + 1CO3 | 5.7 [70] | −23.9 | 1PrOH + 1CO3 | | −22.9 [26] |
| 2PrOH + 2CO2 | 7.9 [70] | −21.2 | 1PrOH + 2CO2 | | −22.1 [26] |
| 2BuOH + 1CO1 | 7.5 [33] | −24.8 | 1BuOH + 1CO1 | | −25.5 [26] |
| 2BuOH + 1CO2 | 6.2 [80] | −24.5 | 1BuOH + 1CO2 | | −25.4 [26] |
| 2PrOH + ACP | 8.0 [97] | −24.4 | 1PrOH + ACP | | −25.2 [82] |
| 2BuOH + ACP | 8.3 [97] | −24.2 | 1BuOH + ACP | | −24.6 [82] |
| 2PrOH + DPE | 7.3 [47] | −16.7 | 1PrOH + DPE | 6.5 [47] | −17.5 |
| 2PrOH + DBE | 7.2 [71] | −16.5 | 1PrOH + DBE | | −16.4 [53] |
| 2BuOH + DPE | 8.0 [31] | −16.0 | 1BuOH + DPE | 4.5 [105] | −19.5 |
| 2BuOH + DBE | 7.9 [81] | −15.8 | 1BuOH + DBE | 7.9 [122] | −15.9 |
| 2PrOH + THF | 5.4 [51] | −21.0 | 1PrOH + THF | 4.5 [51] | −21.8 |

TABLE 4 (continued)

| | | | | | |
|---|---|---|---|---|---|
| 2PrOH + THP | 6.0 [51] | −19.4 | 1PrOH + THP | 4.5 [51] | −20.9 |
| 2PrOH + 1,4DX | 8.9 [51] | −21.6 | 1PrOH + 1,4DX | 7.2 [52] | −23.3 |
| 2BuOH + THF | 5.0 [88] | −21.4 | 1BuOH + THF | 4.4 [88] | −22.0 |
| 2BuOH + 1,4-DX | 8.4 [89] | −22.1 | 1BuOH + 1,4-DX | 7.9 [52] | −22.6 |
| 2BuOH + 1,3DX | 10.3 [89] | −20.5 | 1BuOH + 1,3DX | 9.6 [123] | −21.2 |
| 2PrOH + DMC | 9.9 [126] | −22.8 | 1PrOH + DMC | | −25.4 [27] |
| 2BuOH + DMC | 13.2 [126] | −19.5 | 1BuOH + DMC | | −25.3 [27] |
| 2BuOH + DEC | 11.2 [126] | −19.1 | 1BuOH + DEC | | −19.0 [27] |

[a]for symbols, see Table 2.

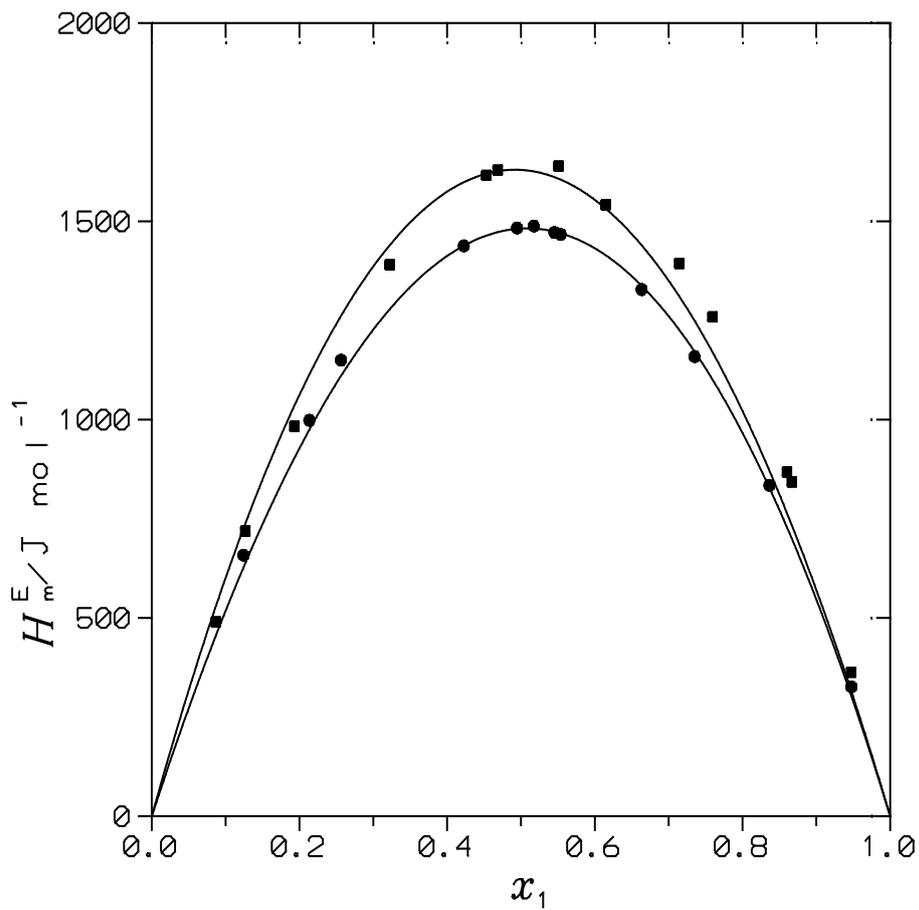

**Figure 1**     $H_m^E$ for 2-alkanol (1) + 2-butanone (2) systems at 298.15 K. Points, experimental results: (●), 2-propanol [79]; (■), 2-butanol [80]. Solid lines, Flory calculations with interaction parameters listed in Table 2.

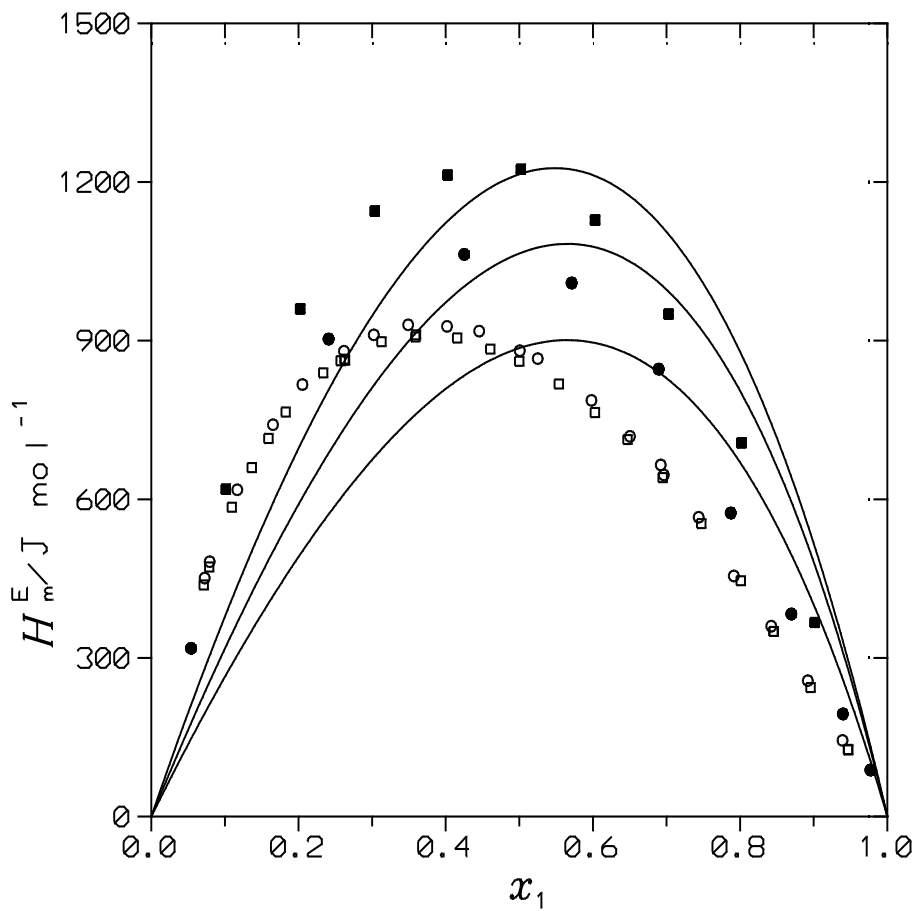

**Figure 2**  $H_m^E$ for alkanol (1) + dibutylether (2) systems at 298.15 K. Points, experimental results: (●), 2-propanol [71]; (■), 2-butanol [81]; (O), 1-propanol, (□), 1-butanol [122]. Solid lines, Flory calculations with interaction parameters listed in Table 2.

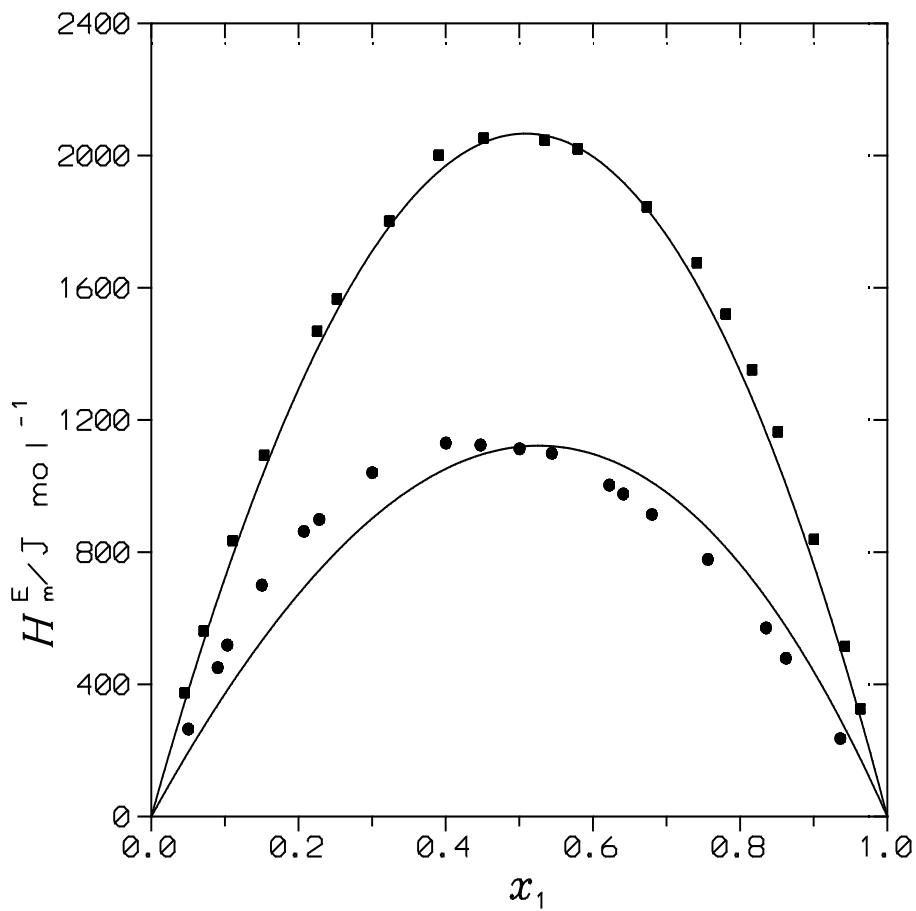

**Figure 3**    $H_m^E$ for 2-propanol (1) + cyclic ether (2) systems at 298.15 K. Points, experimental results [51]: (●), oxolane; (■), 1,4-dioxane. Solid lines, Flory calculations with interaction parameters listed in Table 2.

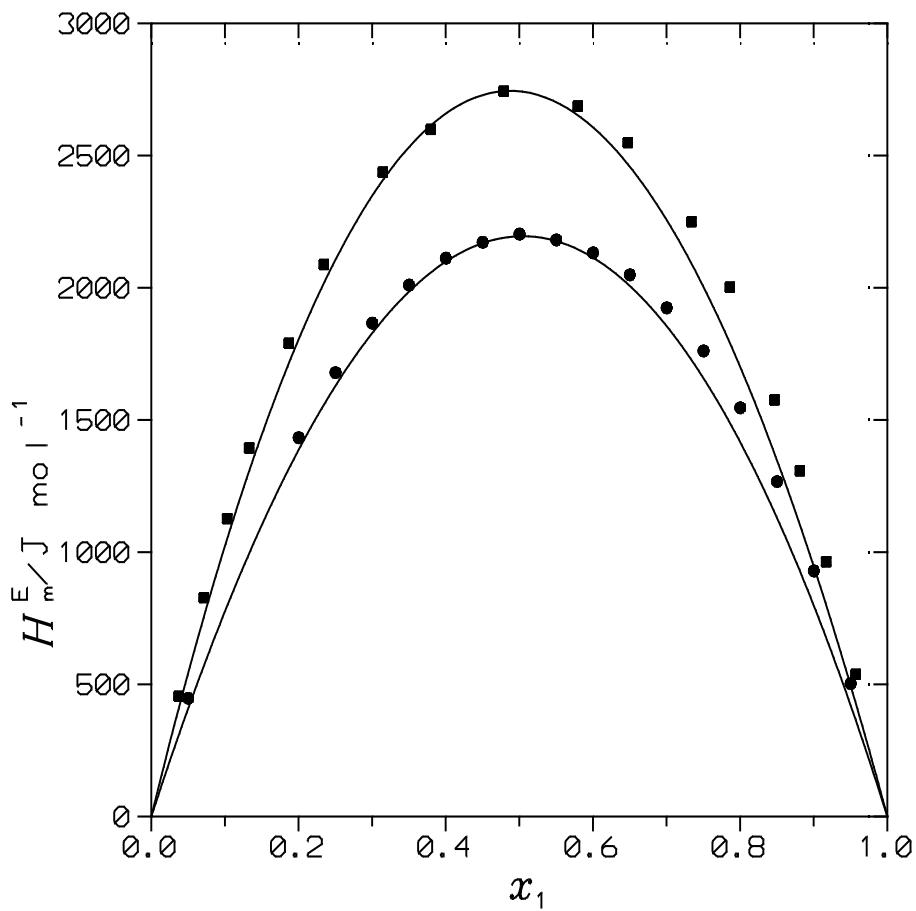

**Figure 4**  $H_m^E$ for 2-alkanol (1) + dimethyl carbonate (2) systems at 298.15 K. Points, experimental results: (●), 2-propanol [98]; (■), 2-butanol [99]. Solid lines, Flory calculations with interaction parameters listed in Table 2.

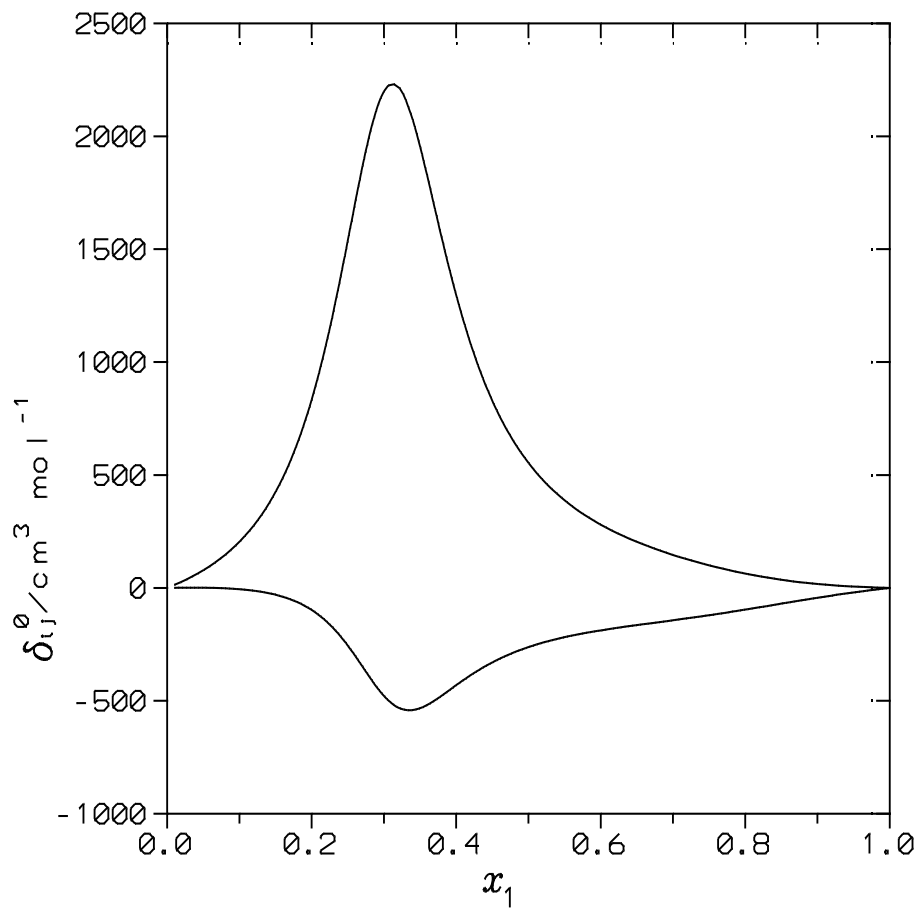

**Figure 5**   Linear coefficients of preferential solvation, $\delta_{ij}^0$, for the 2-propanol (1) + heptane (2) system at 298.15 K. Upper curve, $\delta_{11}^0$; lower curve, $\delta_{12}^0$ [92,131].

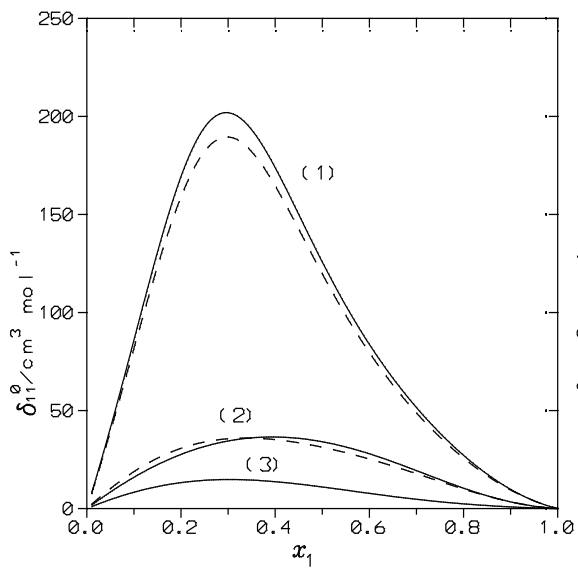 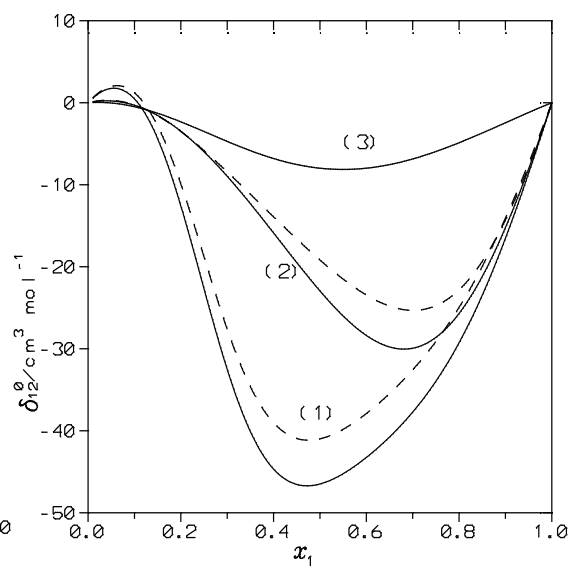

**Figure 6a**  **Figure 6b**

**Figure 6**     Linear coefficients of preferential solvation, $\delta_{ij}^0$, for the alkanol (1) + organic solvent (2) system at 298.15 K. Solid lines, 2-propanol mixtures, dashed lines, 1-propanol solutions: (1), dipropyl ether; (2), 2-butanone; (3), oxolane (see Table 3 for source of data)

# THERMODYNAMICS OF 2-ALKANOL + POLAR ORGANIC SOLVENT MIXTURES. I. SYSTEMS WITH KETONES, ETHERS OR ORGANIC CARBONATES

Juan Antonio González,[*] Fernando Hevia, Luis Felipe Sanz, Daniel Lozano Martín and Isaías. García de la Fuente

[a]G.E.T.E.F., Departamento de Física Aplicada, Facultad de Ciencias, Universidad de Valladolid, Paseo de Belén, 7, 47011 Valladolid, Spain.

*corresponding author, e-mail: jagl@termo.uva.es; Fax: +34-983-423136; Tel: +34-983-423757

**TABLE S1**

Excess molar volumes, $V_m^E$, variation of $V_m^E$ with temperature $(\frac{\Delta V_m^E}{\Delta T})_p$ and isochoric molar excess internal energies, $U_{Vm}^E$, at equimolar composition and 298.15 K for 2-alkanol (1) + organic solvent (2) systems.

| System[a] | $V_m^E$ / cm³ mol⁻¹ | | $(\frac{\Delta V_m^E}{\Delta T})_p$ / cm³ mol⁻¹ K⁻¹ | $U_{Vm}^E$ /J mol⁻¹ |
|---|---|---|---|---|
| | Exp.[b] | Flory[c] | | |
| 2PrOH + 1CO1 | 0.290 [s1] | 1.752 | | 1528 |
| 2PrOH + 1CO2 | 0.219 [79] | 1.482 | 3.9 10⁻³ [79,s2] | 1420 |
| 2PrOH + 1CO3 | 0.230 [s3] | 1.389 | | 1433 |
| 2PrOH + 2CO2 | 0.181 [s3] | 1.343 | | 1409 |
| 2BuOH + 1CO1 | 0.399 [s4] | 1.806 | − 2.6 10⁻⁴ [69, s4] | 1738 |
| 2BuOH + 1CO2 | 0.352 [80] | 1.492 | 1.2 10⁻³ [80,s2] | 1517 |
| 2PrOH + ACP | − 0.247 [46] | 0.857 | | 1646 |
| 2BuOH + ACP | 0.077 [46] | 1.418 | | 2235 |
| 2PrOH + DPE | − 0.027 [47] | 0.929 | | 969 |
| 2PrOH + DBE | 0.133 [71] | 0.981 | | 1028 |
| 2BuOH + DPE | − 0.073 [31] | 0.861 | | 1089 |
| 2BuOH + DBE | 0.120 [81] | 1.022 | | 1178 |
| 2PrOH + THF | 0.267 [127] | 1.057 | | 1027 |
| 2PrOH + THP | 0.177 [s5] | 1.136 | 3.1 10⁻³ [s5] | 1072 |
| 2PrOH + 14DX | 0.255 [s6] | 1.544 | 4.3 10⁻³ [s6] | 1978 |
| 2PrOH + 13DX | 0.204 [s7] | | 6.2 10⁻³ [s7] | |
| 2BuOH + THF | 0.298 [107] | 1.056 | | 1042 |
| 2BuOH + THP | 0.285 [107] | 1.713 | | |
| 2BuOH + 14DX | 0.510 [s8] | 1.639 | 2.4 10⁻³ [s8,s9] | 1985/1891 |
| 2BuOH + 13DX | 0.433 [s9] | 1.846 | 2.7 10⁻³ [s9] | 2178 |
| 2PrOH + DMC | 0.478 [s10] | 1.985 | 2.0 10⁻³ [s10] | 2031 |
| 2PrOH + DEC | 0.403 [108] | | 5.1 10⁻³[108] | |
| 2BuOH + DMC | 0.749 [s10] | 2.402 | 1.65 10⁻³[s10] | 2474 |
| 2BuOH + DEC | 0.594 [108] | 2.363 | 1.3 10⁻³ [108] | 2078 |

**TABLE S1 (continued)**

| | | | |
|---|---|---|---|
| 2PrOH + PC | −0.060 [126] | 0.982 | 2396 |
| 2BuOH + PC | 0.190 [126] | 1.082 | 2336 |

[a]for symbols, see Table 2; [b]experimental result; [c]value calculated with interaction parameters listed in Table 2.

**TABLE S2**

Excess molar enthalpies, $H_m^E$, at equimolar composition and temperature $T$ for 2-alkanol (1) + organic solvent(2) mixtures. Values of the Flory interaction parameter, $X_{12}$, and of the relative standard deviations for $H_m^E$, $\sigma_r(H_m^E)$, are also included

| System[a] | $N$[b] | $T$/K | $X_{12}$ /J cm$^{-3}$ | $H_m^E$/ J mol$^{-1}$ | $\sigma_r(H_m^E)$ [c] | Ref |
|---|---|---|---|---|---|---|
| 2PrOH + 1CO1 | 19 | 343.15 | 91.42 | 1468 | 0.022 | 95 |
| 2PrOH + 1CO1 | 19 | 363.15 | 87.38 | 1426 | 0.040 | 95 |
| 2PrOH + ACP | 12 | 348.15 | 112.26 | 2077 | 0.054 | 97 |
| 2PrOH + DMC | 19 | 328.15 | 150.29 | 2527 | 0.045 | 98 |
| 2BuOH + DEC | 17 | 318.15 | 121.50 | 2574 | 0.039 | 100 |

[a]for symbols, see Table 2; [b]number of data points; [c]$\sigma_r(H_m^E) = \left[ \dfrac{1}{N} \sum \left( \dfrac{H_{m,exp}^E - H_{m,calc}^E}{H_{m,exp}^E} \right)^2 \right]^{1/2}$

**References**

s1. M Iglesias, B. Orge, J. Tojo. Densities, refractive indices, and derived excess properties of {$x_1$CH$_3$COCH$_3$ + $x_2$CH$_3$OH + (1- $x_1$- $x_2$)CH$_3$CH(OH)CH$_3$} at the temperature 298.15 K. J. Chem. Thermodyn. 27 (1995) 1161-1167.

s2. M. Almasi. Densities and viscosities of binary mixtures of ethylmethylketone and 2-alkanols; application of the ERAS model and cubic EOS. Thermochim. Acta 554 (2013) 25-31.

s3. T.M. Letcher, J.A. Nevines. Excess volumes of (ketone + alkanol) at the temperature 298.15 K. J. Chem. Eng. Data 40 (1995) 293-295.

s4. B. Orge, M. Iglesias, G. Marino, J. Tojo. Thermodynamic properties of (acetone + methanol + 2-butanol) at $T$ = 298.15 K. J. Chem. Thermodyn. 31 (1999) 497-512.

s5. Y.C. Kao, C.H. Tu. Densities, viscosities, refractive indices, and surface tensions for binary and ternary mixtures of 2-propanol, tetrahydropyran, and 2,2,4-trimethylpentane. J. Chem. Thermodyn. 43 (2011) 216-226.

s6. M. Contreras. Densities and viscosities of binary mixtures of 1,4-dioxane with 1-propanol and 2-propanol at (25,30,35, and 40) °C. J. Chem. Eng. Data 46 (2011) 1149-1152.

s7. H.C. Ku, C.C. Wang, C.H. Tu. Densities, viscosities, refractive índices, and surface tensions for mixtures of 1,3-dioxolane + 2-propanol or + 2,2,4-trimethylpentane at (288.15. 298.15 and 308.15) K and 1,3-dioxolane + 2-propanol + 2,2,4-trimethylpentane at 298.15 K. J. Chem. Eng. Data 54 (2009) 131-136

s8. A.G. Camacho, M.A. Postigo, G.C. Pedrosa, I.L. Acevedo, M. Katz. Densities, refractive índices and excess properties of mixing of the *n*-octanol + 1,4-dioxane + 2-butanol ternary system at 298.15 K. Can. J. Chem. 78 (2000) 1121-1127.

s9. I. Gascón, S. Martín, P. Cea, M.C. López, F.M. Royo. Density and speed of sound for binary mixtures of a cyclic ether with a butanol isomer. J. Solution Chem. 31 (2002) 905-915.

s10. A. Rodríguez, J. Canosa, J. Tojo. Density, refractive index, and speed of sound of binary mixtures (diethyl carbonate + alcohol) at several temperatures. J. Chem. Eng. Data 46 (2001) 1476-1486.

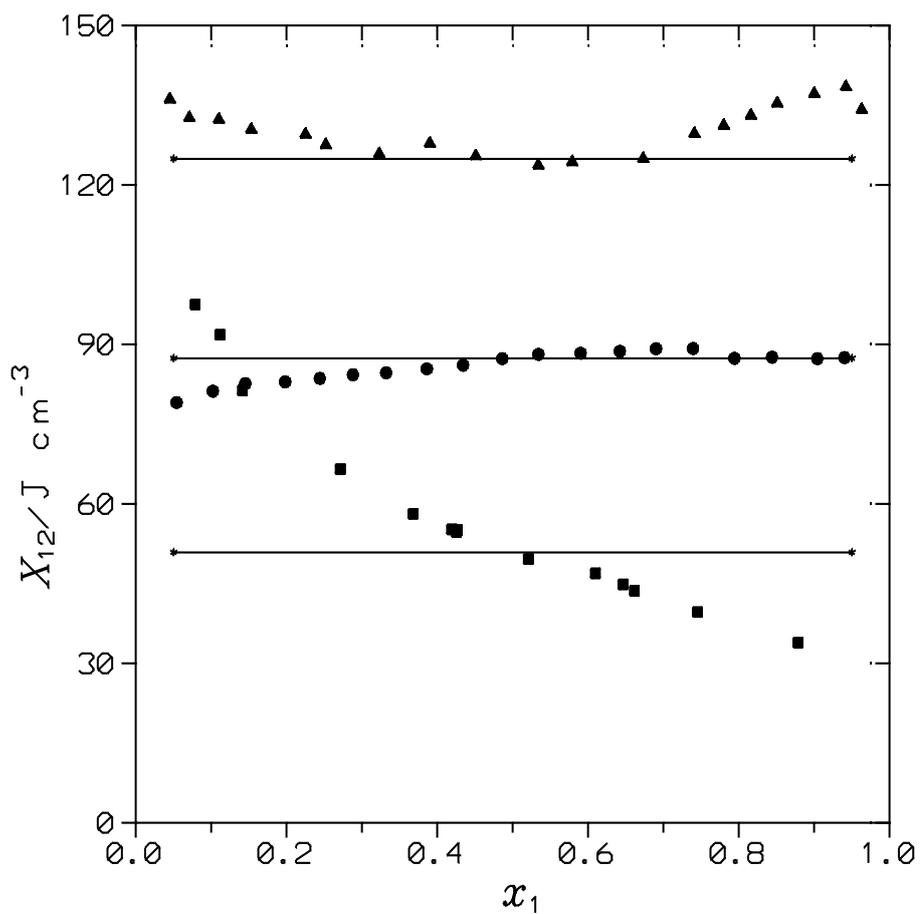

**Figure S1.** Flory interaction parameters, $X_{12}$, for systems containing 2-propanol: (■), dipropyl ether ($T$ = 298.15 K [47]); (●), propanone ($T$ = 363.15 K [95]); (▲), 1,4-dioxane ($T$ = 298.15 K [51]). Solid lines are the corresponding $X_{12}$ values at equimolar composition (Tables 2 and S2).

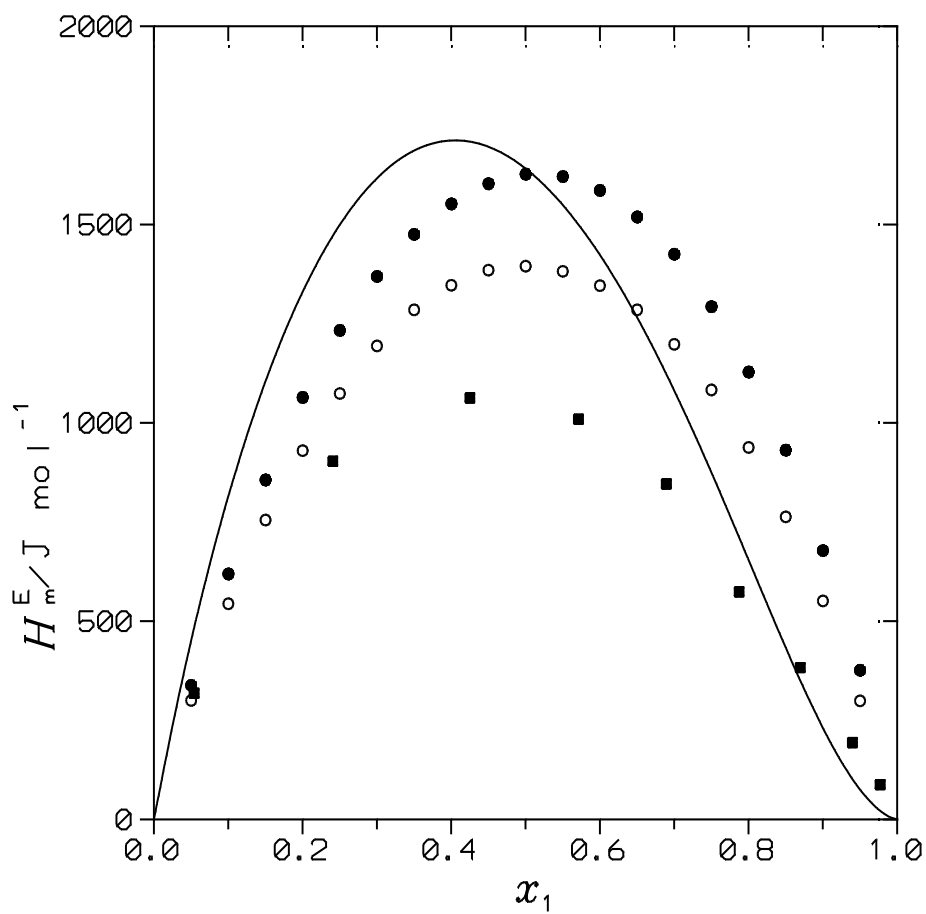

**Figure S2.** $H_m^E$ for the 2-propanol (1) + propanone (2) system at 298.15 K; (●), experimental results [32]; solid line, ERAS calculations with interaction parameters given in text. Experimental values for the mixtures: 2-propanol (1) + dibutyl ether (2) (■) [71], and 1-propanol (1) + propanone (2) (O) [118] at 298.15 K are also given for comparison.